\documentclass[11pt, a4paper]{article}

\pdfoutput=1
\usepackage{jheppub}
\usepackage[utf8]{inputenc}
\usepackage{epsfig}
\usepackage{graphicx}
\usepackage[multiple]{footmisc}
\usepackage{amssymb,amsmath,mathrsfs}
\usepackage{fancybox,framed,tikz}
\usetikzlibrary{decorations.pathmorphing,patterns}
\usepackage{dsfont}
\usepackage{mathtools}
\usepackage{braket}
\usepackage{slashed}
\usepackage{rotating}
\usepackage{bbold,amsfonts}
\usepackage{multirow}
\usepackage{verbatim}
\usepackage{xcolor}

\newcommand{\beq}{\begin{equation}}
\newcommand{\eeq}{\end{equation}}

\renewcommand{\a}{\alpha}
\renewcommand{\b}{\beta}

\renewcommand{\d}{\delta}
\newcommand{\pa}{\partial}
\newcommand{\g}{\gamma}

\newcommand{\D}{\Delta}
\newcommand{\e}{\epsilon}

\renewcommand{\l}{\lambda}

\newcommand{\s}{\sigma}

\renewcommand{\t}{\tau}

\newcommand{\Tr}{\textup{Tr}}
\newcommand{\Str}{\textup{Str}}

\author[a,b]{Lorenzo Bianchi,}
\author[c,d]{Michelangelo Preti,}
\author[e]{Edoardo Vescovi}

\affiliation[a]{Institut f\"ur Theoretische Physik, Universit\"at Hamburg\\
Luruper Chaussee 149, 22761 Hamburg, Germany}

\affiliation[b]{Center for Research in String Theory - School of Physics and Astronomy Queen Mary\\
University of London, Mile End Road, London E1 4NS, UK}

\affiliation[c]{Laboratoire de Physique Th\'eorique, D\'epartement de Physique de l'ENS, \'Ecole Normale Sup\'erieure\\
rue Lhomond 75005 Paris, France}

\affiliation[d]{PSL Universit\'es, Sorbonne Universit\'es, CNRS}

\affiliation[e]{Institute of Physics, University of S\~ao Paulo\\
05314-070 S\~ao Paulo, Brazil}

\emailAdd{lorenzo.bianchi@qmul.ac.uk}
\emailAdd{michelangelo.preti@lpt.ens.fr}
\emailAdd{vescovi@if.usp.br}

\abstract{
In this paper we study the Bremsstrahlung functions for the $\frac16$BPS and the $\frac12$BPS Wilson lines in ABJM theory. First we use a superconformal defect approach to prove a conjectured relation between the Bremsstrahlung functions associated to the geometric ($B^{\varphi}_{1/6}$) and R-symmetry ($B^{\theta}_{1/6}$) deformations of the $\frac16$BPS Wilson line. This result, non-trivially following from a defect supersymmetric Ward identity, provides an exact expression for $B^{\theta}_{1/6}$ based on a known result for $B^{\varphi}_{1/6}$. Subsequently, we explore the consequences of this relation for the $\frac12$BPS Wilson line and, using the localization result for the multiply wound Wilson loop, we provide an exact closed form for the corresponding Bremsstrahlung function. Interestingly, for the comparison with integrability, this expression appears particularly natural in terms of the conjectured interpolating function $h(\lambda)$. During the derivation of these results we analyze the protected defect supermultiplets associated to the broken symmetries, including their two- and three-point correlators.}

\title{Exact Bremsstrahlung functions in ABJM theory}

\begin{document}
\maketitle

\section{Introduction and results}
Exact results for interacting quantum field theories are notoriously hard to achieve. Recent years, however, have seen spectacular developments in the computation of exact physical observables for conformal field theories with extended supersymmetry. These theories, despite their little phenomenological interest, constitute an important laboratory for testing our understanding of the finite coupling regime  of quantum field theories. 
\bigskip

The maximally supersymmetric theory in four dimensions, $\mathcal{N}=4$ Super Yang-Mills (SYM), is a celebrated example where supersymmetric localization \cite{Pestun:2016zxk}  as well as the discovery of an integrable structure \cite{Minahan:2002ve} made such developments possible. While the former applies to a restricted class of protected observables (those that are annihilated by some supercharges), the latter proved very powerful for the computation of planar anomalous dimensions, inherently unprotected quantities. Therefore, despite the application of integrability has recently been extended to a wider range of observables, for some time it has been difficult to find a physical quantity accessible to both techniques. Luckily, the authors of \cite{Correa:2012at} realized that the energy emitted by a moving particle, commonly known as Bremsstrahlung function, is a good candidate. On the one hand, it is suited for the integrability approach \cite{Correa:2012hh,Drukker:2012de} since it appears in the small angle expansion of the cusp anomalous dimension. On the other hand, interpreting the Wilson line as a superconformal defect \cite{Correa:2012at,Cooke:2017qgm,Giombi:2017cqn}, the Bremsstrahlung function can be related to the first-order deformation of the circular Wilson loop expectation value \cite{Fiol:2012sg,Correa:2012at}, known exactly via localization \cite{Pestun:2009nn,Erickson:2000af,Drukker:2000rr,Pestun:2007rz}. The same match between integrability \cite{Gromov:2013qga,Gromov:2015dfa} and localization \cite{Bonini:2015fng}  happens for the generalized Bremsstrahlung function with $L$ units of R-charge.  
\bigskip

Besides providing a highly non-trivial check of the result, computing the same quantity in two different ways allows to understand the precise identification of the parameters. Indeed, every integrability computation features a parameter $h$ whose relation with the 't Hooft coupling $\l$ cannot be fixed by symmetry considerations. Whereas such relation turns out to be trivial for $\mathcal{N}=4$ SYM, this is not the case for its three-dimensional relative $\mathcal{N}=6$ super Chern-Simons theory with matter, known as ABJM theory \cite{Aharony:2008ug}. In the latter case, a conjectured expression for $h(\l)$ \cite{Gromov:2014eha} agrees with weak \cite{Gaiotto:2008cg,Grignani:2008is,Nishioka:2008gz,Minahan:2009aq,Minahan:2009wg,Leoni:2010tb} and strong \cite{McLoughlin:2008he, Abbott:2010yb,LopezArcos:2012gb,Bianchi:2014ada} coupling perturbative computations (see also \cite{Cavaglia:2016ide} for the generalization to ABJ theory), but an exact derivation is still missing.
\bigskip

Similarities between ABJM and $\mathcal{N}=4$ SYM include the existence of a known string theory dual, the emergence of an integrable structure and the high degree of supersymmetry (although it is not maximal for ABJM). A crucial difference, instead, is the preserved supersymmetry of Wilson line operators. While the $\mathcal{N}=4$ SYM Maldacena-Wilson loop \cite{Maldacena:1998im} preserves half of the supercharges (thus denoted as $\frac12$BPS), its obvious ABJM generalization is annihilated by only $1/6$ of the supercharges ($\frac16$BPS) \cite{Berenstein:2008dc,Drukker:2008zx,Chen:2008bp}. A $\frac12$BPS Wilson loop for ABJM, whose existence was expected as the dual to the fundamental string on $AdS_4\times \mathbb{CP}^3$, was built in  \cite{Drukker:2009hy} by introducing local couplings to the fermionic fields in the gauge superconnection. 
\bigskip

The bosonic and fermionic Wilson loops are $\frac16$BPS and $\frac12$BPS respectively when their contour is maximally symmetric, i.e. a straight line or a circle. A smooth deformation of the contour, if combined with a suitable modification of the gauge connection, may still preserve a fraction of the original supersymmetry \cite{Cardinali:2012ru}. On the other hand, when the Wilson line is deformed by a cusp, the supersymmetry is completely broken and the expectation value diverges. The coefficient of the divergent term, whose form can be analyzed in very general terms  \cite{Polyakov:1980ca,Korchemsky:1987wg}, is called cusp anomalous dimension. Miming the four-dimensional case \cite{Drukker:1999zq,Drukker:2011za}, one can introduce a second deformation by an internal angle $\theta$ entering the local couplings with the bosonic and fermionic fields in the gauge superconnection. In this case the generalized cusp anomalous dimension $\Gamma_{\text{cusp}}(\varphi, \theta)$ would depend on both angles.
\bigskip

For ABJM, two different generalized cusps may be defined for the bosonic and fermionic Wilson lines \cite{Griguolo:2012iq,Correa:2014aga}. While for the former no residual BPS configuration could be found, for the latter the specific case $\varphi^2=\theta^2$ still preserves two supercharges, such that $\Gamma^{1/2}_{\text{cusp}}(\varphi, \pm\varphi)$ vanishes. This particular feature has interesting consequences for the small angle expansions of the two cusp anomalous dimensions. For the fermionic case one has
\begin{equation}\label{brem12}
 \Gamma^{1/2}_{\text{cusp}}(\varphi, \theta) \sim (\theta^2-\varphi^2) B_{1/2}
\end{equation}
which is the analogue of the four-dimensional case. This fact, supported by a three-loop computation, led to the conjecture of a relation between the Bremsstrahlung function and the first-order supersymmetric deformation of the circular $\frac12$BPS Wilson loop (often denoted as latitude Wilson loop\footnote{This nomenclature, which we follow here, may be misleading since moving a Wilson loop from the equator to a parallel on a sphere corresponds, through a conformal mapping, to a dilatation on the plane, thus not affecting its expectation value. Nevertheless, to preserve supersymmetry, the deformation of the contour is accompanied by a modification of the superconnection which leads to a non-trivial dependence of the expectation value on the deformation parameter.}) \cite{Bianchi:2014laa,Bianchi:2017svd}. This relation was finally proven in \cite{Bianchi:2017ozk} by relating $B_{1/2}$ to a particular combination of bosonic and fermionic two-point functions inserted on the Wilson line. In presence of a localization result for the latitude circular Wilson loop, this property would allow to compute $B_{1/2}$ exactly. Unfortunately no matrix model representation for the latitude Wilson loop is available and one has to rely on the fact that the bosonic and fermionic Wilson loops are cohomologically equivalent, i.e. their difference is exact with respect to a combination of the preserved supercharges. Elaborating on this and making some further assumptions,  $B_{1/2}$ can be expressed in terms of the complex phase appearing in front of the Wilson loop expectation value, when computed with a contour splitting regularization \cite{Bianchi:2014laa,Bianchi:2016yzj}. 
\bigskip

For the bosonic generalized cusp one can define two different Bremsstrahlung functions 
\begin{equation}\label{brem16}
 \Gamma^{1/6}_{\text{cusp}}(\varphi, \theta) \sim \theta^2 B^{\theta}_{1/6}-\varphi^2B^{\varphi}_{1/6}\,.
\end{equation}
In this case, only $B^{\theta}_{1/6}$ can be related to the first order deformation of a circular Wilson loop \cite{Correa:2014aga}. On the other hand, a completely different argument led to express $B^{\varphi}_{1/6}$ as the first order expansion of a $n$-wound Wilson loop for $n\to1$ \cite{Lewkowycz:2013laa}. Given the availability of localization results for the $n$-wound circular Wilson loop \cite{Kapustin:2009kz,Marino:2009jd,Drukker:2010nc,Klemm:2012ii} $B^{\varphi}_{1/6}$ is known exactly. Recently, the simple relation
\begin{equation}\label{result}
B^{\varphi}_{1/6}=2 B^{\theta}_{1/6}
\end{equation}
was conjectured based on a finite $N$ four-loop computation \cite{Bianchi:2017afp,Bianchi:2017ujp}.
Before this work, the apparent simplicity of \eqref{result} was not backed up by any (even speculative) field-theory argument and the lack of a genuine string computation of $B^{\theta}_{1/6}$ prevented a strong-coupling check, in contrast to the other Bremsstrahlung functions in ABJM.

\bigskip

In this paper we show that the identification \eqref{result} is a consequence of a supersymmetric Ward identity. The latter is derived in the framework of superconformal defects. The $\frac16$BPS Wilson loop preserves a $su(1,1|1)\oplus su(2) \oplus su(2)$ subalgebra of the original $osp(6|4)$ ABJM symmetry algebra. The residual symmetry can be used to constrain defect correlation functions of local operators inserted along the Wilson line. Such insertions are organized in irreducible representations of the preserved subalgebra: long multiplets, whose scaling dimension is not protected, and short multiplets which are annihilated by one of the two preserved supercharges and whose dimension is fixed by algebraic arguments. Among the latter, we are particularly interested in those associated to the broken symmetries. Whenever a defect breaks a symmetry of the original theory the conservation law for the associated currents should be supplemented by some defect degrees of freedom. This relation with the previously conserved currents guarantees that these defect excitations are protected. A well studied example is the displacement operator which compensates for the non-conservation of the stress tensor and accounts for the breaking of translation invariance \cite{Billo:2016cpy}. For supersymmetric theories we know that the stress tensor belongs to the same supermultiplet of the supersymmetry and R-symmetry currents \cite{Cordova:2016emh}. For the defect setup we will clarify, using some algebraic arguments and showing explicit expressions, that the displacement operator is the top component (as should be expected \cite{Liendo:2016ymz}) of a supermultiplet containing also a fermionic operator associated to some broken supercharges. The defect excitations associated to the rest of the broken supercharges together with the R-symmetry ones form a different supermultiplet, which we denote as $R$-multiplet.
\bigskip 

Focusing on these two supermultiplets we explore the constraints of the residual symmetries on two- and three-point functions. This task is eased by the observation that the preserved superalgebra coincides with the chiral part or the $\mathcal{N}=2$ superconformal algebra in two dimensions (of course only the global part, not the infinite dimensional super-Virasoro extension). This allows to exploit the results of \cite{Blumenhagen:1992sa,DiVecchia:1985ief,Mussardo:1988av} for correlation functions in superspace\footnote{Notice that this coincidence also opens the way to a conformal bootstrap approach to the study of operator insertions on the Wilson line. In this context the superconformal blocks derived in \cite{Cornagliotto:2017dup,Fitzpatrick:2014oza} should be suitable also for the defect field theory and it would be interesting to study similarities and differences between the defect and the full two-dimensional SCFT.}. The outcome of this analysis is that two-point functions of operators belonging to the displacement and the R-multiplet are completely fixed by superconformal symmetry up to an overall factor, their Zamolodchikov norm. Interestingly, the Zamolodchikov norm of the displacement operator is related to the Bremsstrahlung function $B^{\varphi}_{1/6}$ \cite{Correa:2012at}, while that of the R-multiplet is proportional to $B^{\theta}_{1/6}$ \cite{Correa:2014aga}. Therefore the relation \eqref{result} establishes a connection between the Zamolodchikov norms of the two supermultiplets. We show that this connection can be derived through an unconventional Ward identity which uses the action of a supercharge that is not preserved by the defect. Such Ward identity, as it usually happens for defect field theories, mixes two- and three-point functions, but in our case we will be able to show that the symmetry is large enough to set to zero all the involved three-point functions, thus allowing to prove \eqref{result}.
\bigskip

After this derivation we explore the consequences of our result. First of all we point out how equation \eqref{result} leads to a useful ansatz for the relation between the winding number $n$ and the deformation parameter $\nu$, characterizing the deformation of the maximal circular Wilson loop. We then extend this prescription to a relation between winding and framing and this allows to rederive the exact form for $B_{1/2}$ conjectured in \cite{Bianchi:2014laa} as well as a relation between $B_{1/2}$ and $B_{1/6}^{\varphi}$. We conclude our work by expressing $B_{1/2}$ in a closed form. This may result useful from the point of view of integrability, since, if we take for granted the conjecture of \cite{Gromov:2014eha} for $h(\lambda)$, our expression involves elliptic functions whose argument is naturally expressed in terms of $h$ rather than $\lambda$.
\bigskip

This paper is organized as follows. In section \ref{sec:2} we review the properties of bosonic and fermionic supersymmetric Wilson loops in ABJM, in particular we focus on the cusped Wilson loops and the related cusp anomalies and Bremsstrahlung functions. We also introduce some background for the study of Wilson line excitations as a defect one-dimensional superconformal field theory. In section \ref{symmetry} we study this field theory and the structure of their protected supermultiplets under the symmetry preserved by the bosonic line. In section \ref{sec:4} we introduce the relevant two- and three-point functions in both the defect theory and the related superspace. Exploiting some Ward identities involving conserved and non-conserved supercharges, we relate the Zamolodchikov norms of certain two-point functions to compute $B^\theta_{1/6}$ exactly. In section \ref{sec:5}  we establish a relation between all the Bremsstrahlung functions via a unique function. The main result of section \ref{matrix_model} is the exact closed form of $B_{1/2}$ in terms of the conjectured interpolating function $h(\lambda)$. Few appendices follow, which contain conventions and some details of the supermultiplets and the supersymmetry algebra.

\section{BPS Wilson loops and the Bremsstrahlung functions in ABJM}\label{sec:2}

The $\mathcal{N}=6$ Chern-Simons-matter theory, known as ABJM \cite{Aharony:2008ug,Bandres:2008ry}, is a three-dimensional superconformal field theory with $U(N)_{k}\times U(N)_{-k}$ gauge group, with $k$ being the integer Chern-Simons level. Its global symmetry is $OSp(6|4)$: the bosonic subsector of the supergroup contains the R-symmetry group $SO(6)\sim SU(4)$ and the Euclidean 
conformal group in three-dimensions $Sp(4)\sim SO(1,4)$, the fermionic subsector generates the $\mathcal{N}=6$ supersymmetries.

The theory has the following field content: two gauge fields ${(A_\mu)_i}^j$ and 
${(\hat{A}_\mu)_{\hat{i}}}^{\hat{j}}$, belonging respectively to
the adjoint of $U(N)_{k}$ and $U(N)_{-k}$, four complex scalars
${(C_I)_i}^{\hat{j}}$ (or ${(\bar{C}^I)_{\hat{i}}}^j$) as well as four complex fermions ${(\bar{\psi}^I)_i}^{\hat{j}}$ (or ${(\psi_{I})_{\hat{i}}}^j$) belonging to the bifundamental (antibifundamental) of the gauge group $U(N)_{k}\times U(N)_{-k}$. 

In this paper we are mostly interested in supersymmetric Wilson loops, a rich class of BPS observables that, in principle, can be known exactly. 
We start by reviewing the definition and properties of the bosonic and fermionic Wilson loop operators. We focus on a particular choice for the contour (cusped Wilson line) in order to define the cusp anomalous dimensions and the Bremsstrahlung functions summarizing the state of the art in the literature. 

\subsection{The bosonic Wilson loop}\label{sec:2.1}

The bosonic Wilson loop \cite{Berenstein:2008dc,Drukker:2008zx,Chen:2008bp,Rey:2008bh} is a natural generalization of the four-dimensional Wilson-Maldacena loop \cite{Maldacena:1998im} and it is defined as follows
\begin{equation}\label{eq:WL}
\mathcal{W}_B[\mathcal{C}] = \frac{1}{N} \Tr \left[\mathcal{P}\exp\left({ -i \int_{\mathcal{C}} d\tau~ \mathcal{L}_B(\tau)} \right)\right]\quad \text{with}\quad\mathcal{L}_B=A_{\mu} \dot x^{\mu}-\frac{2 \pi i}{k} |\dot x|\, M_{J}^{\ \ I} C_{I}\bar C^{J}\,,
\end{equation}
where $\mathcal{C}$ is the path along which the loop is supported parametrized by $x(\tau)$, $\mathcal{P}$ is the path-ordering operator and the trace $\Tr$ is taken in the fundamental representation of $U(N)_k$. One can also define an analogous Wilson loop operator $\hat{\mathcal{W}}_B[\mathcal{C}] $ belonging to the fundamental representation of $U(N)_{-k}$, where the connection $\hat{\mathcal{L}}_B$ contains $\hat{A}_{\mu}$ instead of $A_{\mu}$ and $\hat{M}_{J}^{\ \ I} \bar C^{J}C_{I}$ instead of $M_{J}^{\ \ I} C_{I}\bar C^{J}$. The scalar coupling $M_{J}^{\ \ I}$ and $\hat{M}_{J}^{\ \ I}$ are in general matrices with arbitrary entries. They can be constrained by supersymmetry imposing the standard vanishing condition 
\begin{equation}\label{susyB}
\delta_{\text{susy}}\mathcal{L}_B=0\,.
\end{equation}
and using the supersymmetry transformations in \eqref{N6susy}.
The choice of which supercharges are preserved by the Wilson loop \eqref{eq:WL} fixes both the scalar couplings and the contour parametrization. 

In order to study the most general class of supersymmetric Wilson loops, it is convenient to consider operators lying on curves on the sphere $S^2$. These loops can be mapped to their equivalent operators in flat space through a conformal map that maintains the number of preserved supercharges. In this setting, an arbitrary curve on $S^2$ preserves 1/24 of the total number of supersymmetries and the equator corresponds to the maximally supersymmetric operator, namely $\frac16$BPS. It is also possible to consider an operator with an intermediate number of preserved supersymmetries. This operator is called $\frac{1}{12}$BPS bosonic latitude Wilson loop and it can be written as a two-parameter deformation of the $\frac16$BPS Wilson loop. 
These parameters combine in a single quantity that the Wilson loop depends on \cite{Bianchi:2014laa}
\begin{equation}\label{param}
\nu=\sin{2\alpha}\cos\theta_0\qquad\text{with}\quad -\pi/2\leq\theta_0\leq\pi/2\;\;\text{and}\;\; 0\leq\alpha\leq\pi/4\,.
\end{equation}
The only relevant deformation appears in the scalar couplings of \eqref{eq:WL} which can be written in terms of $\nu$ as follows
\begin{equation}\label{Mbosoniclat}
 {M_J}^I(\nu)={\hat{M}_J}{}^I(\nu)=\begin{pmatrix} -\nu & 0 & e^{-i\tau}\sqrt{1-\nu^2}  & 0\\
          0 & -1 & 0 & 0\\
          e^{i\tau}\sqrt{1-\nu^2} & 0 & \nu & 0\\
          0 & 0 & 0 & 1
         \end{pmatrix}\,.
\end{equation}
The expectation value of the latitude Wilson loop depends only on $\nu$, so we refer to the operator as $\mathcal{W}_B(\nu)$. 

In the limit $\nu\rightarrow 1$ we recover the $\frac16$BPS Wilson loop on the maximal circle of $S^2$. Through a particular conformal mapping, we can project this loop on the plane obtaining the $\frac16$BPS infinite straight Wilson line with the contour parametrized by
\begin{equation}\label{paramline}
x^{\mu}=\{\tau,0,0\}\qquad\text{with}\quad -\infty < \tau < \infty
\end{equation}
and the scalar couplings given by
\begin{equation}\label{Mbosonic}
 {M_J}^I(1)={\hat{M}_J}{}^I(1)=\begin{pmatrix} -1 & 0 & 0 & 0\\
          0 & -1 & 0 & 0\\
          0 & 0 & 1 & 0\\
          0 & 0 & 0 & 1
         \end{pmatrix}\,.
\end{equation}
This operator preserves a $su(1,1|1)\oplus su(2) \oplus su(2)$ subalgebra of $osp(6|4)$ (see appendix \ref{subalgebra} for the details).
In the following we refer to it as $\mathcal{W}_{1/6}$. We summarize the relations between the supersymmetric Wilson loops above as follows
\begin{equation}
\underset{\text{$\frac{1}{12}$BPS latitude on $S^2$}}{\mathcal{W}_B(\nu)}\xrightarrow{\nu \rightarrow 1}\underset{\text{$\frac16$BPS circle on $S^2$}}{\mathcal{W}_B(1)}\xrightarrow{\text{conf. map}}\underset{\text{$\frac16$BPS line in $\mathbb{R}^3$}}{\mathcal{W}_{1/6}}\,.
\end{equation}

A string configuration for the bosonic Wilson loop is still elusive. Since the fundamental string ending along the Wilson loop contour on the boundary of $AdS_4$ and localized in $\mathbb{CP}^3$ preserves more supercharges, a smearing of the string over a $\mathbb{CP}^1$ is expected to break the supersymmetry in order to match with the gauge theory observable \cite{Drukker:2008zx,Rey:2008bh}.

\subsection{The fermionic Wilson loop}

In order to match the number of supercharges preserved by the fundamental string in $AdS_4\times\mathbb{CP}^3$, on the field theory side one needs to consider the holonomy of a  $U(N|N)$ superconnection \cite{Drukker:2009hy,Lee:2010hk}. This operator with an arbitrary contour was given in \cite{Cardinali:2012ru} where it was expressed as
\begin{equation}\label{WLF}
 \mathcal{W}_F[\mathcal{C}]=\frac{1}{\Str\, \mathcal{T}}\,\Str\left[\mathcal{P}\exp\left(-i\int_{\mathcal{C}} d\tau \,\mathcal{L}_F(\t) \right) \mathcal{T}\right]
\end{equation}
with a superconnection $\mathcal{L}_F(\t)$
\begin{equation}\label{superconnection}
\mathcal{L}_F=\begin{pmatrix} A_\mu \dot{x}^\mu-\frac{2\pi i}{k} |\dot x| {\mathcal{M}_J}^I C_I \bar C^J & -i\sqrt{\frac{2\pi}{k}}|\dot{x}|\eta_I \bar \psi^I\\
             -i\sqrt{\frac{2\pi}{k}}|\dot{x}| \psi_I \bar \eta^I & \hat A_\mu \dot x^\mu-\frac{2\pi i}{k} |\dot x| {\hat{\mathcal{M}}_J}^I  \bar C^J C_I
            \end{pmatrix}\,.
\end{equation}
Here $\Str$ stands for the usual supertrace taken in the fundamental representation and the quantities ${\mathcal{M}_J}^I(\tau)$, ${\hat{\mathcal{M}}_J}^I(\tau)$, $\eta_I(\tau)$ and $\bar \eta^I(\tau)$ are local couplings. As for the bosonic case, one can determine the form of the couplings in terms of the contour $x^{\mu}(\tau)$ by the requirement of preserving some of the supercharges. In this case the standard vanishing condition $\d_\text{susy}\mathcal{L}_F=0$ is too strong and it can be replaced by the weaker requirement \cite{Drukker:2009hy,Lee:2010hk,Cardinali:2012ru}
\begin{equation}\label{susyvarL}
\d_\text{susy}\mathcal{L}_F=\mathcal{D}_{\tau} \mathcal{G}\equiv\pa_{\tau} \mathcal{G}+i[\mathcal{L},\mathcal{G}] \,,
\end{equation}
where $\mathcal{G}$ is a $u(N|N)$ supermatrix. The twist supermatrix $\mathcal{T}$ in \eqref{WLF} is needed for the operator to be gauge invariant.

We are interested in some particular configurations of the fermionic Wilson loop: the latitude and maximal circle on $S^2$ and the infinite straight line in $\mathbb{R}^3$. Using the relation \eqref{susyvarL} and the supersymmetry transformations \eqref{N6susy}, it turns out that the fermionic maximal circle is $\frac12$BPS, matching the supersymmetry preserved by the fundamental string in $AdS_4\times\mathbb{CP}^3$. As in the previous case the latitude Wilson loop can be seen as a two-parameter deformation of the maximally supersymmetric operator on the equator. Those parameters can be rearranged again in the single quantity \eqref{param} and the Wilson loop expectation value depends only on it. We then refer to the 1/6 fermionic latitude as $\mathcal{W}_F(\nu)$. The operator on the maximal circle can be recovered in the limit $\nu\rightarrow 1$. Using the stereographic conformal projection, this operator is mapped in the $\frac12$BPS fermionic Wilson line lying on the contour parametrized by \eqref{paramline}. This operator preserves a $su(1,1|3)$ subalgebra of $osp(6|4)$ (see \cite{Bianchi:2017ozk} for the details). In the following we refer to it as $\mathcal{W}_{1/2}$. All the details about the scalar and fermionic couplings for any of the previous configurations can be found in \cite{Drukker:2009hy,Cardinali:2012ru}. We repeat  below the relation among fermionic Wilson loops for clarity:
\begin{equation}
\underset{\text{$\frac16$BPS latitude on $S^2$}}{\mathcal{W}_F(\nu)}\xrightarrow{\nu \rightarrow 1}\underset{\text{$\frac12$BPS circle on $S^2$}}{\mathcal{W}_F(1)}\xrightarrow{\text{conf. map}}\underset{\text{$\frac12$BPS line in $\mathbb{R}^3$}}{\mathcal{W}_{1/2}}\,.
\end{equation}

\subsection{The generalized cusp and the Bremsstrahlung functions in ABJM}\label{gencusp}

\begin{figure}[htbp]
  \begin{center}
  \includegraphics[width=10.5cm]{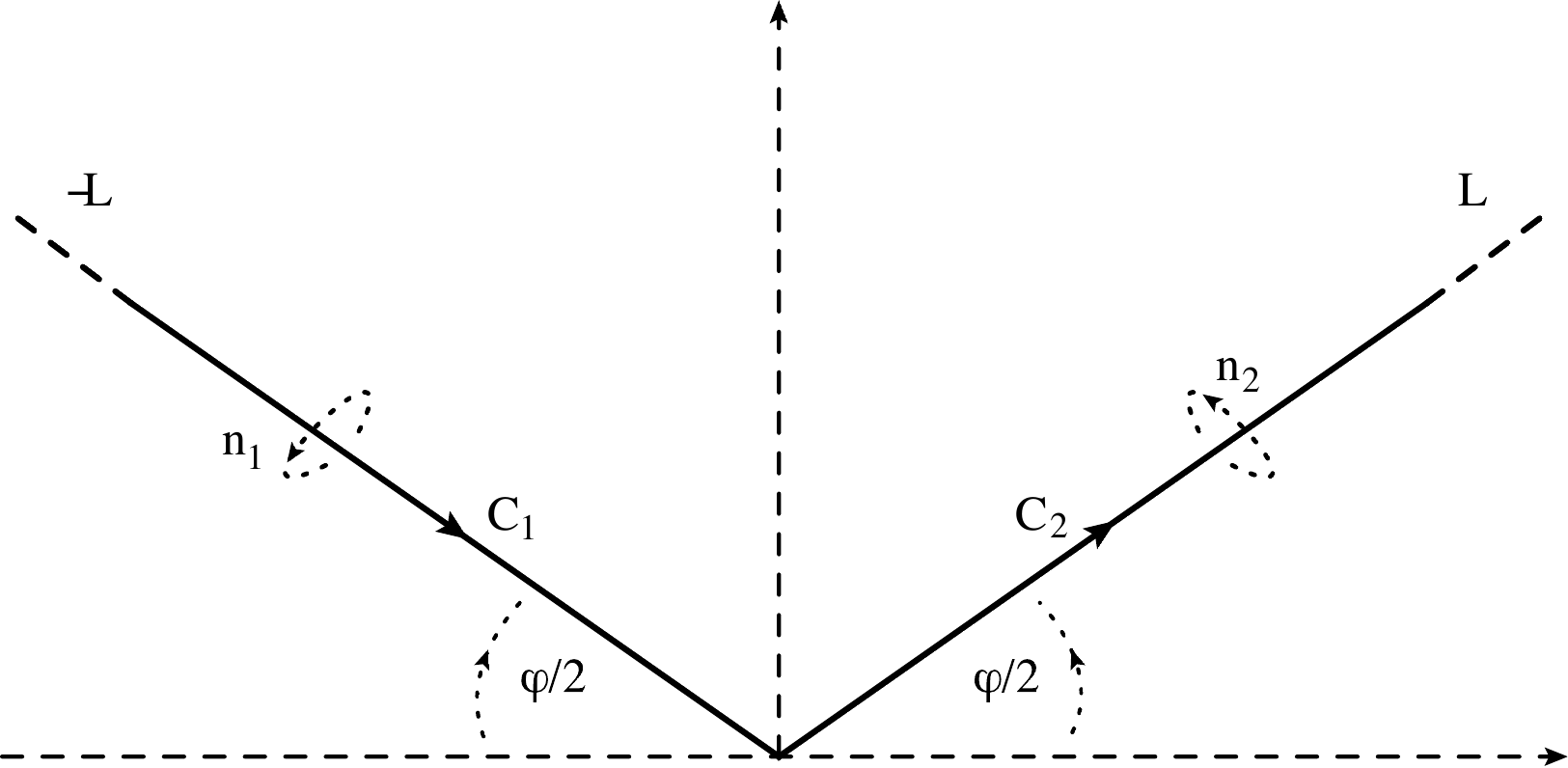}
  \caption{The planar Euclidean cusp with angular opening $\pi-\varphi$ between the Wilson lines.}
  \label{fig:cusp}
  \end{center}
\end{figure}

Let us start consider a bosonic or fermionic Wilson line with contours $C_1$ and $C_2$ in $\mathbb{R}^3$ intersecting in the origin and forming the curve $C=C_1\cup C_2$ (see Figure \ref{fig:cusp}) with parametrization
\begin{equation}\label{cuspcontour}
C:\qquad x^\mu=\{\tau\cos\frac\varphi 2,|\tau|\sin\frac\varphi 2,0\}\qquad\text{with}\quad -L\leq\tau\leq L\,,
\end{equation}
where $L$ is an IR cut-off. In general one can introduce an extra parameter $\theta$ that corresponds to the angular separation of the Wilson lines on $C_1$ and $C_2$ in the R-symmetry space. 

This deformation affect the scalar and fermionic couplings in \eqref{eq:WL} and \eqref{WLF}. Considering the factorized form of the fermionic coupling $\eta_I^\alpha=n_I\eta^\alpha$ (the same for the complex conjugate) and that $\hat{M}=M$ and $\hat{\mathcal{M}}=\mathcal{M}$, the contractions\footnote{The contractions are defined as $M_i\cdot M_j={(M_i)_J}^I{(M_j)_I}^J$ and $n_i\cdot \bar{n}_j= (n_i)_I(\bar{n}_j)^I$.} of the couplings are defined as follows
\begin{equation}
M_i\cdot M_j = \mathcal{M}_i\cdot\mathcal{M}_j=
     \begin{cases}
       4 \cos^2\frac{\theta}{2} &\quad i\neq j\\
       4                                   &\quad i=j
     \end{cases}\qquad
     n_i\cdot \bar{n}_j =
     \begin{cases}
       2i \cos\frac{\theta}{2} &\quad i\neq j\\
       2i                                   &\quad i=j
     \end{cases}
     \end{equation}
where $i,j=1,2$ and $\{M_1,\mathcal{M}_1,n_1,\bar{n}_1\}\in C_1$ and $\{M_2,\mathcal{M}_2,n_2,\bar{n}_2\}\in C_2$. The twist supermatrix $\mathcal{T}$ in \eqref{WLF} in this case is
\begin{equation}
\mathcal{T}= \begin{pmatrix}
              \mathbb{1}_N 	& 0\\
              0						&-\mathbb{1}_N 
             \end{pmatrix}
\end{equation}
and the supertrace becomes the usual trace in the fundamental representation and the normalization coefficient $\Str\,\mathcal{T}=2N$. This configuration is called generalized cusp.

Unlike the infinite straight lines, the bosonic and fermionic generalized cusped Wilson lines do not preserve any of the supersymmetries and develop logarithmically divergent, which lead in turn to the definition of the associated anomalous dimensions as
\begin{equation}\begin{split}\label{Gammacusp}
 &\log \braket{\mathcal{W}_B[C]}\sim-\Gamma_{\text{cusp}}^{1/6}(\theta,\varphi) \log\frac{L}{\e}+\mbox{finite}\\\
  &\log \braket{\mathcal{W}_F[C]}\sim-\Gamma_{\text{cusp}}^{1/2}(\theta,\varphi) \log\frac{L}{\e}+\mbox{finite}
\end{split}\end{equation}
where $C$ is the cusp contour \eqref{cuspcontour}, $L$ and $\epsilon$ are an IR and UV regulator respectively.
The coefficients of the logarithms $\Gamma_{\text{cusp}}^{1/6}$ and $\Gamma_{\text{cusp}}^{1/2}$ are the cusp anomalous dimensions of the bosonic and fermionic cusped Wilson lines respectively. They are two important  physical observables that control the IR divergences of scattering amplitudes of massive colored particles, besides the important properties outlined in the introduction. Furthermore, setting $\theta=0$ and performing the analytic continuation $\varphi\rightarrow i\infty$, one finds the light-like cusp anomalous dimension, whose value is computed exactly via integrability \cite{Beisert:2006ez,Gromov:2008qe}.

Over the last years, $\Gamma_{\text{cusp}}^{1/6}$ was studied at weak coupling in \cite{Griguolo:2012iq} (for the strong coupling see the final comment of section \ref{sec:2.1}). Also $\Gamma_{\text{cusp}}^{1/2}$ was extensively studied at both weak and strong coupling. Its value was computed at two loops via perturbation theory \cite{Griguolo:2012iq} and exactly in the double scaling limit where only ladder diagrams contribute ($\lambda\rightarrow 0$, $i\theta\rightarrow\infty$ and $\lambda e^{i\theta/2}=\text{const}$) \cite{Bonini:2016fnc}. The case $\varphi=0$ was explored at three loops using the HQET formalism in \cite{Bianchi:2017svd, Preti:2017fjb}. On the string theory side $\Gamma_{\text{cusp}}^{1/2}$ is known at next to leading order from \cite{Forini:2012bb,Correa:2014aga}.

As mentioned in the introduction, see \eqref{brem12} and \eqref{brem16}, the small angle limit of the cusp anomalous dimension gives the Bremsstrahlung functions. The exact computation of the Bremsstrahlung functions in ABJM is an arduous task for which a complete proof is still elusive. Nevertheless, there exist some conjectured relations that connect each Bremsstrahlung function to the expectation value of the multiply-wound circular Wilson loop with bosonic couplings
\begin{equation}\label{WLBn}
\mathcal{W}^n_B(1)= \frac{1}{N} \Tr \left[\mathcal{P}\exp\left({ -i \oint_0^{2\pi n} d\tau~ \mathcal{L}_B(\tau)} \right)\right]
\end{equation}
where the connection $\mathcal{L}_B(\tau)$ is defined in \eqref{eq:WL}. The new parameter $n$ specifies the number of times that the loop, spanned by $\tau$, wraps the circular contour.
This operator localizes on a matrix model \cite{Kapustin:2009kz} that was solved in detail \cite{Marino:2009jd,Drukker:2010nc,Klemm:2012ii}. The perturbative expansion of this exact result produces predictions for the Bremsstrahlung functions that match the direct computations of the generalized cusps with $\frac12$BPS and $\frac16$BPS rays at both weak and strong coupling.

A formula for the Bremsstrahlung function $B_{1/2}$, defined in \eqref{brem12} as the small angle limit of the cusp with $\frac12$BPS rays, was proposed in \cite{Bianchi:2014laa} as the derivative of the fermionic Wilson loop evaluated on a latitude on $S^2$ with respect to the deformation parameter $\nu$
\begin{equation}\label{B_half_1}
B_{1/2}=\frac{1}{4\pi^2} \Big. \partial_\nu \log \langle \mathcal{W}_F (\nu) \rangle \Big|_{\nu=1}\,.
\end{equation}
The relation was modelled upon an analogous result in $\mathcal{N} = 4$ SYM \cite{Correa:2012at} and it was later proven by applying superconformal defect constraints on the $\frac12$BPS Wilson line insertions \cite{Bianchi:2017ozk}. The relation with the multiply-wound Wilson loop on a circle articulates in a few steps that are reviewed in section \ref{sec:5}. 

The prediction for the Bremsstrahlung function $B^{\varphi}_{1/6}$ associated to the geometric cusp formed with $\frac16$BPS rays was formulated in \cite{Lewkowycz:2013laa}
\begin{equation}\label{B_sixth_varphi}
B^{\varphi}_{1/6}=\frac{1}{4\pi^2} \Big. \partial_n \log |\langle\mathcal{W}^{n}_B\rangle| \Big|_{n=1}\,,
\end{equation}
where the expectation value is computed at framing 1 (see section \ref{sec:framing}). 
This result stems from a chain of relations between different observables which leads to the $\frac16$BPS Wilson loop that winds around the circle $n$ times, although a few steps are not proven with full rigour\footnote{In particular the authors of \cite{Lewkowycz:2013laa} argued for a simple relation between $B$ and $h$, the constant characterizing the stress tensor one-point function, which also led \cite{Fiol:2015spa} to formulate a conjecture for the exact Bremsstrahlung function in $\mathcal{N}=2$ superconformal theories in 4d. Such a relation is not universal and the conditions for its validity in the general framework of defect CFTs have not been clarified yet (see \cite{Billo:2016cpy,Bianchi:2015liz,Bianchi:2016xvf,Lemos:2017vnx,Balakrishnan:2017bjg,Dong:2016wcf,Balakrishnan:2016ttg,Herzog:2017xha} for recent discussions).}. It is not necessary to review them in this paper and one can refer to \cite{Bianchi:2014laa} for a concise summary. However, it is useful to remember that  \eqref{B_sixth_varphi} passed a non-trivial test for the first few weak-coupling perturbative orders \cite{Griguolo:2012iq,Bianchi:2014laa} and it is consistent with string-theory calculations at leading and subleading order \cite{Correa:2014aga, Aguilera-Damia:2014bqa}.

In this paper we focus on the Bremsstrahlung function $B^{\theta}_{1/6}$ that corresponds to a cusp distortion in R-symmetry space along a $\frac16$BPS straight line. A relation with the bosonic loop~\footnote{The presence of the absolute value is discussed and motivated in appendix \ref{app:correa}.}
\begin{equation}\label{B_sixth_theta_1}
B^{\theta}_{1/6}=\frac{1}{4\pi^2} \Big. \partial_\nu \log |\langle \mathcal{W}_B (\nu) \rangle | \Big|_{\nu=1}
\end{equation}
was derived by means of a similar proof in $\mathcal{N}=4$ SYM \cite{Correa:2012at} which involves two-point function of scalar operators inserted along the circular Wilson loop \cite{Correa:2014aga}. The notable complication with respect to the four-dimensional case is again the fact that the right-hand side in \eqref{B_sixth_theta_1} is only known perturbatively, thus preventing the derivation of an all-loop expression for $B^{\theta}_{1/6}$. A few perturbative orders were checked from the bosonic Wilson loop at two loops \cite{Bianchi:2014laa}. At the moment a strong-coupling check is hindered by the lack of the string configuration dual to the latitude Wilson loop with bosonic couplings, despite some attempts of  ``string smearing" \cite{Correa:2014aga}.
There exists a proposed relation for $B^{\theta}_{1/6}$ given in \eqref{result} that relates it to the putative \eqref{B_sixth_varphi}.

\subsection{Wilson lines as superconformal defects}
In this paper we are interested in computing the expectation value of a Wilson loop with local operators inserted along the contour. Given some local operators $\mathcal{O}_i(\tau_i)$, one can define the gauge invariant Wilson line with insertions
\begin{equation}\label{WOOO}
\mathcal{W}[\mathcal{O}_1(\tau_1)\mathcal{O}_2(\tau_2)\,...\,\mathcal{O}_n(\tau_n)]\equiv 
\Tr \mathcal{P} \left[\mathcal{W}_{\tau_i,\tau_1}\mathcal{O}_1(\tau_1)\mathcal{W}_{\tau_1,\tau_2}\mathcal{O}_2(\tau_2)\,...\,\mathcal{O}_n(\tau_n)\mathcal{W}_{\tau_n,\tau_f}\right]
\end{equation}
where $\mathcal{W}_{\tau_a,\tau_b}$ is a fermionic or bosonic Wilson line that starts at position $x(\tau_a)$ and ends at position $x(\tau_b)$. Also, $\tau_i=\tau_f$ for a closed loop and $\tau_i=-\infty$ and $\tau_f=\infty$ for an infinite straight line. Since the local operators $\mathcal{O}_i(\tau_i)$ are inserted between (untraced) Wilson lines, they have to transform in the same representation of the gauge group.
In this paper we are only interested in operator insertions on the bosonic Wilson loop $\mathcal{W}_B$, whose connection transforms in the adjoint of $U(N)_{k}$, so we look at operators belonging to the same representation.
The vacuum expectation value of \eqref{WOOO} can be interpreted as a $n$-point correlation function of local operators where the vacuum is the supersymmetric Wilson loop $\mathcal{W}$:
\begin{equation}\label{npoint}
\langle\mathcal{O}_1(\tau_1)\mathcal{O}_2(\tau_2)\,...\,\mathcal{O}_n(\tau_n)\rangle_{\mathcal{W}}\equiv
\frac{\langle\mathcal{W}[\mathcal{O}_1(\tau_1)\mathcal{O}_2(\tau_2)\,...\,\mathcal{O}_n(\tau_n)]\rangle}{\langle\mathcal{W}\rangle}\,.
\end{equation}
This is nothing but a correlation function in a one-dimensional defect. When the residual symmetry preserved by the corresponding Wilson loop includes the conformal group, the correlation functions defined by \eqref{npoint} satisfy all the axioms of a one-dimensional CFT. For instance, if the contour is a straight line, the fermionic and bosonic defects preserve $SU(1,1|3)$ \cite{Bianchi:2017ozk} and $SU(1,1|1)$ (see section \ref{symmetry}) superconformal groups respectively. In this setting, operator insertions can be organized according to the representations of the preserved supergroup. Particular care, however, must be devoted to the conformal descendants. Indeed for Wilson loop correlators one has the defining property
\begin{equation}\label{defprop}
\langle \,...\, \mathcal{D}_\tau\mathcal{O}(\tau)\,...\,\rangle_{\mathcal{W}}=\partial_\tau\langle\, ...\, \mathcal{O}(\tau)\,...\,\rangle_{\mathcal{W}}
\end{equation}
where $\mathcal{D}_\tau$ is the covariant derivative taken with respect to the total connection of the loop $\mathcal{W}$. For the case of interest here, we consider the following definition
\begin{equation}
\mathcal{D}_\tau\mathcal{O}(\tau)\equiv \partial_\tau+i[\mathcal{L}_B,\mathcal{O}(\tau)]\,.
\end{equation}
The implication of equation \eqref{defprop} is that, in building the representations of the superconformal algebra for inserted operators, the covariant derivative $\mathcal{D}_\tau$ plays the role of the ordinary derivative in standard CFT, i.e.  the generator of translations along the line.

In the following we will need to consider the action of a generic infinitesimal variation on a Wilson line. This  translates into operator insertions as
\begin{equation}\label{insert}
\delta (\log \mathcal{W}[\mathcal{C}])=-i\int_{\mathcal{C}} d\tau\,\langle \delta \mathcal{L}(\tau)\rangle_{\mathcal{W}}
\end{equation}
where $\mathcal{L}$ is the bosonic or fermionic Wilson loop connection depending on which operator $\mathcal{W}$ we are considering. The operator $\delta$ corresponds to any infinitesimal transformation. In particular, if it represents a supersymmetry transformation, it can be written as $\delta=\epsilon \mathcal{Q}$ where $\mathcal{Q}$ is a supercharge and $\epsilon$ a Grassmann infinitesimal parameter. In this paper we are interested in infinitesimal deformations of two-point functions on the defect that can be written as
{\begingroup\makeatletter\def\f@size{10}\check@mathfonts
\begin{equation}\begin{split}\label{vivalatopa}
\delta \langle\mathcal{O}_1(\tau_1)\mathcal{O}_2(\tau_2)\rangle_{\mathcal{W}}\!=& \langle\delta\mathcal{O}_1(\tau_1)\mathcal{O}_2(\tau_2)\rangle_{\mathcal{W}}\!+ \langle\mathcal{O}_1(\tau_1)\delta\mathcal{O}_2(\tau_2)\rangle_{\mathcal{W}}\!-\!i\!\!\int_{\tau_i}^{\tau_1} \!\!\!d\tau\langle \delta \mathcal{L}(\tau)\mathcal{O}_1(\tau_1)\mathcal{O}_2(\tau_2)\rangle_{\mathcal{W}}\!\\
&-i\int_{\tau_1}^{\tau_2} d\tau\,\langle \mathcal{O}_1(\tau_1)\delta \mathcal{L}(\tau)\mathcal{O}_2(\tau_2)\rangle_{\mathcal{W}}-i\int_{\tau_2}^{\tau_f} d\tau\,\langle \mathcal{O}_1(\tau_1)\mathcal{O}_2(\tau_2)\delta \mathcal{L}(\tau)\rangle_{\mathcal{W}}\,.
\end{split}\end{equation}\endgroup}
This transformation generates defect three-point functions when it is applied to the Wilson loop as in \eqref{insert}. Notice that, if the deformation is a supersymmetric transformation, the relative signs in the right-hand side of \eqref{vivalatopa} can change depending on the position of the Grassmann infinitesimal parameter. Also, if the vacuum of the original theory is invariant under the variation $\d$, one can use equation \eqref{vivalatopa} to derive Ward identities as we do in section \eqref{Wardbrok}.

\section{Symmetry considerations}\label{symmetry}
The $\frac16$BPS Wilson line defined in \eqref{eq:WL}, when the contour is an infinite straight line or a circle, preserves a $su(1,1|1)\oplus su(2) \oplus su(2)$ subalgebra of the full $osp(6|4)$ ABJM superalgebra. Commutation relations of such subalgebra are given in appendix \ref{subalgebra}, where we also review its representation theory. Here we consider the fundamental fields of the theory and how they organize in representations of the preserved symmetries.
The $SU(4)$ R-symmetry group is broken down to $SU(2)\times SU(2) \times U(1)$. The preserved supercharges $ Q^{12}_+\equiv Q$ and $ Q^{34}_-\equiv \bar Q$ are neutral under the two $SU(2)$ factors and oppositely charged under $U(1)$ (this $U(1)$ is a bosonic subalgebra of $SU(1,1|1)$ as detailed in appendix \ref{subalgebra}). Therefore the action of the supercharges on a highest weight state does not affect the $SU(2)$ charges, but only its $U(1)$ charge.   Matter fields can be split according to the new symmetry. We use an index $a$ for the fundamental representation of the first $SU(2)$ and an index $\dot a$ for the second one. For the bosons we have
\begin{align}
&& C_I=(C_a&,C_{\dot a}) &  \bar C^I=(\bar C^a&,\bar C^{\dot a}) \\
 U(1)&  \text{ charge} & (-\tfrac12&, \phantom{-} \tfrac12) &  (\phantom{-}\tfrac12&,  -\tfrac12) \nonumber 
\end{align}
whereas for the fermions
\begin{align}
&& \psi^+_I=(\psi^+_a&,\psi^+_{\dot a})& \psi^-_I=(\psi^-_a&,\psi^-_{\dot a}) &  \bar \psi_+^I=(\bar \psi_+^a&,\bar \psi_+^{\dot a}) & \bar \psi_-^I=(\bar \psi_-^a&,\bar \psi_-^{\dot a})\\
 U(1)&  \text{ charge} & (\phantom{-}0&, \phantom{-} 1) &  (-1&,  \phantom{-}0) & (\phantom{-}0&,-1) & (\phantom{-}1&,\phantom{-}0)\nonumber
\end{align}
In the present paper we are interested in operator insertions on the Wilson lines. Those insertions are organized in representations of the preserved superalgebra and we will be concerned with operators which belong to short multiplets. In appendix \ref{subalgebra} we show that the $su(1,1|1)$ superalgebra allows for two possible shortening conditions, leading to the $\frac12$BPS multiplets $\mathcal{B}_j$, annihilated by Q, and  $\bar{\mathcal{B}}_j$, annihilated by $\bar Q$. Looking at the supersymmetry transformations in appendix \ref{pressusy} we immediately find the first examples of short multiplets. Indeed the operators $C_a$ and $\bar C_{\dot a}$ are annihilated by $\bar Q$ and they are superprimaries of  a $\bar{\mathcal{B}}_{-\frac12}$, while the operators $\bar C^a$ and $C_{\dot a}$ are superprimaries of $\mathcal{B}_{\frac12}$.
\begin{center}
\begin{tikzpicture}
  \node[above] at (0,4.8) {$\bar C^{a}$};
  \draw[->]  (-0.35,4.9)--(-0.65,4.6);
  \node[above] at (-1,3.8) {$\e^{ a b} \psi_{ b}^+$};
  \node at (-3,3.5) {$\D$};
  \node at (-2.6,3) {$j$};
  \draw[-] (-2.6,3.4)--(-3,3);
  \node at (0,3) {$\frac12$};
  \node at (-1,3) {$0$};
   \node[above] at (-3,3.8) {$1$};
   \node[above] at (-3,4.8) {$\frac12$};
  \end{tikzpicture}\hspace{1 cm}
\begin{tikzpicture}
  \node[above] at (0,4.8) {$C_{\dot a}$};
  \draw[->]  (-0.35,4.9)--(-0.65,4.6);
  \node[above] at (-1,3.8) {$\e_{\dot a \dot b}\bar \psi^{\dot b}_-$};
  \node at (0,3) {$\frac12$};
  \node at (-1,3) {$0$};
  \end{tikzpicture}\hspace{1.6 cm}
\begin{tikzpicture}
  \node[above] at (0,4.8) {$C_a$};
  \draw[->]  (0.35,4.9)--(0.65,4.6);
  \node[above] at (1,3.8) {$\e_{ab} \bar \psi^b_+$};
  \node at (0,3) {$-\frac12$};
  \node at (1,3) {$0$};
  \end{tikzpicture}\hspace{1 cm}
  \begin{tikzpicture}
  \node[above] at (0,4.8) {$\bar C^{\dot a}$};
  \draw[->]  (0.35,4.9)--(0.65,4.6);
  \node[above] at (1,3.8) {$\e^{\dot a \dot b} \bar \psi_{\dot b}^-$};
  \node at (0,3) {$-\frac12$};
  \node at (1,3) {$0$};
  \end{tikzpicture}
\end{center}
Notice however that these operators change in the bifundamental representation of the gauge group $U(N)_{k}\times U(N)_{-k}$ and in order to build insertions of the kind \eqref{npoint} we would need to use both $\mathcal{W}_B$ and $\hat{\mathcal{W}}_B$. We now look for short multiplets which are singlet under the second $U(N)$ factor, allowing us to use only the Wilson line $\mathcal{W}_B$.

\subsection{Broken currents and defect operators}

The straight Wilson line breaks some of the original symmetries and, as a consequence, some of the currents are no longer conserved. A prototypical example is that of spacetime translation, for which the stress tensor conservation is broken to
\begin{equation}\label{dertmn}
 \pa_{\mu}T^{\mu i}=\d^2(x^j) \mathbb{D}^i(\tau)
\end{equation}
where $i,j=2,3$ are directions orthogonal to the line, the delta function localizes the r.h.s. on the defect profile (a straight line along the direction 1 in this case) and $\mathbb{D}^i$ is the displacement operator. Of course equation \eqref{dertmn} is written in a loose notation and it must be interpreted as a Ward identity when both sides are inserted inside a correlation function with other operators. In the following, we will use complex coordinates in the orthogonal directions and work with the complex combinations $\mathbb{D}=\mathbb{D}_2-i \mathbb{D}_3$ and $\bar{\mathbb{D}}=\mathbb{D}_2+i \mathbb{D}_3$. The broken momentum generators are also organized as $\mathfrak{P}\equiv P_2-iP_3$ and $\bar{\mathfrak P} \equiv P_2+i P_3$. The name displacement operator can be understood by looking at the integrated version of the Ward identity \eqref{dertmn}, which can be schematically written as 
\begin{align}\label{tranlsact}
   \braket{X}_{\d_{\mathfrak{P}}W}=-\int d\t \braket{\mathbb{D}(\tau) X}_{\mathcal{W}} \e + \mathcal{O}(\e^2)& 
\end{align}
where $X$ is an arbitrary set of local operators and the notation $\d_{\mathfrak{P}}W$ means that the profile of the defect is slightly translated in the orthogonal direction conjugate to $\mathfrak{P}$, specifically $\d_{\mathfrak{P}}=\e \mathfrak{P}$. For Wilson lines this formula is particularly convenient since, comparing the identity \eqref{insert} with \eqref{tranlsact} without any operator $X$, we can find an explicit expression for the displacement operator
\begin{equation}
 \mathbb{D}(\tau)=i \mathfrak{P} \mathcal{L}_B
\end{equation}
with $\mathcal{L}_B=A_1+\frac{2\pi i}{k}(C_a \bar C^a-C_{\dot a}\bar C^{\dot a})$.
This reads
\begin{equation}\label{disp}
 \mathbb{D}=i F-\frac{2\pi}{k} D(C_a \bar C^a -C_{\dot a} \bar C^{\dot a})
\end{equation}
where $D$ is the complex combination $D=D_2-i D_3$ of covariant derivatives and $F$ that of field strengths $F=F_{21}-i F_{31}$.

In a supersymmetric theory the stress tensor is not the only current whose conservation is broken. Another bosonic example is the R-symmetry current. Out of the 15 $SU(4)$ generators ${J_I}^J$ (with ${J_I}^I=0$), 7 generate the preserved $SU(2)\times SU(2) \times U(1)$, while the remaining 8 are broken and can be organized as 
\begin{equation}
 \mathfrak{J}^{a \dot a} \qquad \text {and} \qquad \bar{\mathfrak{J}}^{\dot a  a}\,.
\end{equation}
Two sets of currents $j^{\mu a \dot a}$ and $\bar{j}^{\mu \dot a a}$ are no longer conserved and we can write down Ward identities similar to \eqref{dertmn}. In particular (factors of $i$ are inserted for future convenience)
\begin{align}
 \pa_{\mu}j^{\mu a \dot a}=i\d^2(x^j) \mathbb{O}^{a \dot a} \qquad  \pa_{\mu}\bar{j}^{\mu a \dot a}=i\d^2(x^j) \bar{\mathbb{O}}^{\dot a a}\,.
\end{align}
The physical interpretation of $\mathbb{O}^{a\dot a}$ is analogous to that of the displacement operator, but in internal space. Its insertion inside correlation functions accounts for the infinitesimal variation of the Wilson line under a broken R-symmetry generator
 \begin{align}\label{Jact}
   \braket{X}_{\d_{\mathfrak{J}}W}=-i\int d\t \braket{\mathbb{O}^{a \dot a}(\tau) X}_{\mathcal{W}} \e_{a \dot a}  + \mathcal{O}(\e^2)& 
\end{align}
where $\e^{a \dot a}$ is the infinitesimal parameter for the broken R-symmetry transformation such that $\d_{\mathfrak{J}}=\e_{a \dot a} \mathfrak{J}^{a \dot a}$. Once more, exploiting \eqref{insert} we can find the explicit expression of $\mathbb{O}^{a \dot a}$ and its conjugate
\begin{align}\label{OObar}
 \mathbb{O}^{a \dot a}&=\mathfrak{J}^{a \dot a}\mathcal{L}_B=-\frac{4\pi i}{k}\e^{ab} C_{b} \bar C^{\dot a} & \bar{\mathbb{O}}^{\dot a a}&=\bar{\mathfrak{J}}^{a \dot a}\mathcal{L}_B=-\frac{4\pi i}{k}\e^{\dot a \dot b} C_{\dot b} \bar C^{ a}\,.
\end{align}

We now consider the fermionic generators. The set of 10 broken supercharges is organized as follows
\begin{align}
&& \mathfrak{Q}&=Q_+^{34} &  \mathfrak{Q}^{a\dot a}&=\begin{pmatrix} Q_+^{13} & Q_+^{14}\\
                        Q_+^{23} & Q_+^{24}
                       \end{pmatrix} 
 & \bar{\mathfrak{Q}}^{a\dot a}&=\begin{pmatrix} Q_-^{13} &  Q_-^{14} \\
                        Q_-^{23} &  Q_-^{24}
                       \end{pmatrix} & \bar{\mathfrak{Q}}&= Q_-^{12}\\
U(1) &\text{ charge} & &-\tfrac32 & &-\tfrac12 & &\phantom{-}\tfrac12 & &\phantom{-}\tfrac32\\
SU(2)^2 &\text{ repr.} & &[\mathbf{0},\mathbf{0}] & &[\mathbf{2},\mathbf{2}] & &[\mathbf{2},\mathbf{2}] & &[\mathbf{0},\mathbf{0}]
\end{align}
and for their supersymmetry currents $\mathcal{S}^{\mu}$, $\mathcal{S}^{\mu a \dot a}$, $\bar{\mathcal{S}}^{\mu}$ and $\bar{\mathcal{S}}^{\mu a \dot a}$ we have the following Ward identities
\begin{align}
 \pa_{\mu}\mathcal{S}^{\mu}&=i\d^2(x^j) \mathbb{F} & \pa_{\mu}\bar{\mathcal{S}}^{\mu}&=i\d^2(x^j) \bar{\mathbb{F}}\\
  \pa_{\mu}\mathcal{S}^{\mu a \dot a}&=i\d^2(x^j) \mathbb{f}^{a \dot a} & \pa_{\mu}\bar{\mathcal{S}}^{\mu a \dot a}&=i\d^2(x^j) \bar{\mathbb{f}}^{a \dot a}
\end{align}
such that, analogously to the bosonic case
\begin{align}
\mathbb{F}&=\mathfrak{Q} \mathcal{L}_B & \bar{\mathbb{F}}&=\bar{\mathfrak{Q}} \mathcal{L}_B \label{FFbar}\\
\mathbb{f}^{a \dot a}&=\mathfrak{Q}^{a \dot a} \mathcal{L}_B & \bar{\mathbb{f}}^{a \dot a}&=\bar{\mathfrak{Q}}^{a \dot a} \mathcal{L}_B\,.
\end{align}
Using the action of the broken supercharges on the fundamental fields given in appendix \ref{broken} one finds the explicit expressions
\begin{align}
 \mathbb{F}&= \frac{4\pi i}{k}\left(\e^{ab} C_a \psi^-_b + \e_{\dot a \dot b} \bar \psi_+^{\dot a} \bar C^{\dot b} \right) & \bar{\mathbb{F}}&=\frac{4\pi i}{k}\left(\e_{ab} \bar \psi_-^a \bar C^b- \e^{\dot a \dot b} C_{\dot a} \psi^+_{\dot b}\right)\\
 \mathbb{f}^{a \dot a}&= -\frac{4\pi i}{k}\left(\e^{ab} \e^{\dot a \dot b} C_b \psi^-_{\dot b}+\bar \psi^a_+ \bar C^{\dot a}\right) & \bar {\mathbb{f}}^{\dot a  a}&=-\frac{4\pi i}{k}\left(\bar \psi^{\dot a}_- \bar C^{ a} - \e^{\dot a \dot b}\e^{a b} C_{\dot b} \psi_{ b}^+\right)\,.
\end{align}

Notice that all the defect operators introduced in this section, given their relation with previously conserved currents, are protected, i.e. their scaling dimension is fixed to its classical value. This is a clear hint that they should belong to short representations of the preserved superalgebra. On the other hand we also know that in the original theory, ABJM in this case, the stress tensor, the supersymmetry currents and the R-symmetry currents all sit in the same supermultiplet. However no short multiplet of $su(1,1|1)$ could host all the defect degrees of freedom associated to these broken symmetries. Therefore we expect these operators to be arranged in different short multiplets connected by broken supercharges. We will see that this is indeed the case.

\subsection{The displacement supermultiplet}\label{dispmult}
Let us start by considering the commutation relations of the broken supercharges $\mathfrak{Q}$ and $\bar{\mathfrak{Q}}$ with the preserved ones
\begin{align}
 \{Q,\mathfrak Q\}&=2 i \mathfrak{P}  & \{Q,\bar{\mathfrak Q}\}&=0 \label{commpresnotpres1}\\
 \{\bar Q,\bar{\mathfrak Q}\}&=2 i \bar{\mathfrak{P}}  & \{\bar Q,\mathfrak Q\}&=0\,.
\end{align}
 Given these commutation relations we can conclude that the operators $\mathbb{F}$ and $\bar{\mathbb{F}}$
are highest weight operators of two short multiplets since
\begin{align}
 \bar Q \mathfrak{Q} \mathcal{L}_B&=\{\bar Q, \mathfrak{Q}\}\mathcal{L}_B=0  &  Q \bar{\mathfrak{Q}} \mathcal{L}_B&=\{ Q, \bar{\mathfrak{Q}}\}\mathcal{L}_B=0\,.
\end{align}
Furthermore 
\begin{align}
Q\mathbb{F}= Q\mathfrak{Q}\mathcal{L}_B&=\{ Q, \mathfrak{Q}\}\mathcal{L}_B=2 i \mathfrak{P} \mathcal{L}_B & \bar Q\bar{\mathbb{F}}=\bar Q\bar{\mathfrak{Q}}\mathcal{L}_B=\{\bar Q, \bar{\mathfrak{Q}}\}\mathcal{L}_B=2i \bar{\mathfrak{P}} \mathcal{L}_B
\end{align}
which implies
\begin{align}\label{QFD}
 Q \mathbb{F}&=2 \mathbb{D} &  \bar{Q} \bar{\mathbb{F}}&=2 \bar{\mathbb{D}}\,.
\end{align}
Finally, using the commutation relations for the preserved supercharges we have
\begin{align}
 \bar Q\mathbb{D}&=\frac12 \{Q,\bar Q\} \mathbb{F}=-\mathcal{D}_{\tau} \mathbb{F} &  Q\bar{\mathbb{D}}&=\frac12 \{Q,\bar Q\} \bar{\mathbb{F}}=-\mathcal{D}_{\tau} \bar{\mathbb{F}}
\end{align}
in agreement with the general expectation that the displacement operator should be the top component of a short supermultiplet.

In our case we can make everything very explicit. 
Applying a preserved supercharge $Q$ to the expression \eqref{FFbar} for $\mathbb{F}$ we get
\begin{align}
 Q \mathbb{F}=\frac{4\pi}{k} \left(2C_a D\bar C^a-2D C_{\dot a} \bar C^{\dot a}-i\bar \psi^a_+ \psi^-_a-i\bar \psi^{\dot a}_+ \psi^-_{\dot a}\right)\,.
\end{align}
To check that this operator is proportional to the displacement operator one needs to use the equations of motion for the field strength
\begin{equation}
F= \frac{2\pi i}{k} \left(C_aD\bar C^a-DC_a\bar C^a+ C_{\dot a} D\bar C^{\dot a}-DC_{\dot a} \bar C^{\dot a}-i\bar\psi^a_+\psi_a^- -i\bar\psi_+^{\dot a}\psi^-_{\dot a}\right)\,.
\end{equation}
Inserting this in \eqref{disp} gives
\begin{equation}
 \mathbb{D}= \frac{2\pi }{k} \left(2C_a D\bar C^a-2D C_{\dot a} \bar C^{\dot a}-i\bar\psi^a_+\psi_a^- -i\bar\psi_+^{\dot a}\psi^-_{\dot a}\right)
\end{equation}
in agreement with \eqref{QFD}.
We then conclude that $\mathbb{F}$ and $\mathbb{D}$ form a short supermultiplet of type $\bar{\mathcal{B}}_{-\frac32}$. Equivalently $\bar{\mathbb{F}}$ and $\bar{\mathbb{D}}$ sit in a $\mathcal{B}_{\frac32}$ multiplet. 
\begin{center}
\begin{tikzpicture}
  \node[above] at (0,4.8) {$\mathbb{F}$};
  \draw[->]  (0.35,4.8)--(0.65,4.5);
  \node[above] at (1,3.8) {$\mathbb{D}$};
    \node at (0,3) {$-\frac32$};
  \node at (1,3) {$-1$};
  \node at (-1,3.5) {$\D$};
  \node at (-0.6,3) {$j$};
  \draw[-] (-0.6,3.4)--(-1,3);
   \node[above] at (-1,3.8) {$2$};
   \node[above] at (-1,4.8) {$\frac32$};
  \end{tikzpicture}\hspace{1.6 cm}
\begin{tikzpicture}
  \node[above] at (0,4.8) {$\bar{\mathbb{F}}$};
  \draw[->]  (-0.35,4.8)--(-0.65,4.5);
  \node[above] at (-1,3.8) {$\bar{\mathbb{D}}$};
    \node at (0,3) {$\frac32$};
  \node at (-1,3) {$1$};
  \end{tikzpicture}
\end{center}

\subsection{The R-multiplet} \label{scalarmult}
In section \ref{scalarmult} we considered the action of the broken supercharges $\mathfrak{Q}$ and $\bar{\mathfrak{Q}}$ on the connection $\mathcal{L}_B$. Here we consider the action of the other broken supercharges $\mathfrak{Q}^{a\dot a}$. At first sight, the consequences of the commutation relations 
\begin{align}
 \{Q,\mathfrak Q^{a \dot a}\}&=\{Q,\bar{\mathfrak Q}^{a \dot a}\}=0 \label{commpresnotpres2}\\
 \{\bar Q,\mathfrak Q^{a \dot a}\}&=\{\bar Q,\bar{\mathfrak Q}^{a \dot a}\}=0
\end{align}
appear rather puzzling since 
\begin{equation}\label{puzzle}
 Q\mathfrak{Q}^{a\dot a} \mathcal{L}_B=\{ Q,\mathfrak{Q}^{a\dot a}\} \mathcal{L}_B=0 \qquad \qquad \bar Q\mathfrak{Q}^{a\dot a} \mathcal{L}_B=\{ \bar Q,\mathfrak{Q}^{a\dot a}\} \mathcal{L}_B=0\,.
\end{equation}
However one should remember that the r.h.s. of equations \eqref{puzzle} is zero up to gauge transformations, which, for the gauge connection, means a total derivative. In other words the implication of \eqref{puzzle} is that $\mathfrak{Q}^{a \dot a} \mathcal{L}_B$ is the top component of a short multiplet $\bar{\mathcal{B}}_{-1}$. Similarly $\bar{\mathfrak{Q}}^{a \dot a} \mathcal{L}_B$ is the top component of a $\mathcal{B}_{1}$. It is not hard to guess the superprimaries of such multiplets. The operator $\mathbb{O}^{a \dot a}$ ($\bar{\mathbb{O}}^{a \dot a}$) is annihilated by $\bar Q$($Q$)  as one can see from
\begin{align}
 \bar Q \mathfrak{J}^{a \dot a} \mathcal{L}_B&=[\bar Q,\mathfrak{J}^{a \dot a}]\mathcal{L}_B=0 &   Q \bar{\mathfrak{J}}^{\dot a  a} \mathcal{L}_B&=[ Q,\bar{\mathfrak{J}}^{a \dot a}]\mathcal{L}_B=0\,. 
\end{align}
On the other hand 
\begin{align}
 Q \mathbb{O}^{a \dot a}&=[Q,\mathfrak{J}^{a \dot a}]\mathcal{L}_B=\mathfrak{Q}^{a \dot a} \mathcal{L}_B=\mathbb{f}^{a \dot a} & \bar Q \bar{\mathbb{O}}^{\dot a  a}&=[\bar Q,\bar{\mathfrak{J}}^{\dot a a}]\mathcal{L}_B=\bar{\mathfrak{Q}}^{\dot a  a} \mathcal{L}_B=\bar{\mathbb{f}}^{\dot a  a}\,.
\end{align}

As before, this can be checked by explicit computation applying the preserved supercharges on the expressions \eqref{OObar}. We conclude that the operator $\mathbb{O}^{a \dot a}$ ($\bar{\mathbb{O}}^{\dot a a}$) is the superprimary of a $\bar{\mathcal{B}}_{-1}$ ($\mathcal{B}_1$) multiplet.
\begin{center}
\begin{tikzpicture}
  \node[above] at (0,4.8) {$\mathbb{O}^{a\dot a}$};
  \draw[->]  (0.35,4.8)--(0.65,4.5);
  \node[above] at (1,3.8) {$\mathbb{f}^{a \dot a}$};
    \node at (0,3) {$-1$};
  \node at (1,3) {$-\frac12$};
  \node at (-1,3.5) {$\D$};
  \node at (-0.6,3) {$j$};
  \draw[-] (-0.6,3.4)--(-1,3);
   \node[above] at (-1,3.8) {$\frac32$};
   \node[above] at (-1,4.8) {$1$};
  \end{tikzpicture}\hspace{1.6 cm}
\begin{tikzpicture}
  \node[above] at (0,4.8) {$\bar{\mathbb{O}}^{\a \dot a}$};
  \draw[->]  (-0.35,4.8)--(-0.65,4.5);
  \node[above] at (-1,3.8) {$\bar{\mathbb{f}}^{a \dot a}$};
    \node at (0,3) {$1$};
  \node at (-1,3) {$\frac12$};
  \end{tikzpicture}
\end{center}

\section{Correlation functions}\label{sec:4}

The preserved supersymmetry of the $\frac16$BPS Wilson line constraints the form of the correlation functions. Here we explore two different ways to impose those constraints on two- and three-point functions. First we derive Ward idenities for the two-point functions of operators in the R-multiplet and displacement multiplet introduced in sections \ref{scalarmult} and \ref{dispmult} respectively. Afterwards, we rederive the same results and extend them to three-point functions using the more general framework of the superspace. For this, we will be able to exploit some known results based on the coincidence that the preserved superalgebra $su(1,1|1)$ is the same of the holomorphic part of the $\mathcal{N}=2$ supersymmetric theory in two dimensions. We conclude this section with the main result of the paper, i.e. a simple relation between the coefficients $c_s$ and $C_D$ which corresponds precisely to the identity \eqref{result}.

\subsection{Ward identities with preserved supercharges}\label{Wardpres}
We start from operators in the R-multiplet (see section \ref{scalarmult}). The kinematics and R-symmetry structures of the two-point functions are fixed by symmetry
\begin{align}
 \braket{{\mathbb{O}}^{a\dot a}(\tau) \bar{\mathbb{O}}^{\dot b b}(0)}_{\mathcal{W}}&=c_s \frac{\e^{ab} \e^{\dot a\dot b}}{|\tau|^2} &
 \braket{{\mathbb{f}}^{a\dot a}(\tau) \bar{\mathbb{f}}^{\dot b b}(0)}_{\mathcal{W}}= c_f \frac{\e^{ab} \e^{\dot a\dot b} \tau }{|\tau|^4}\,. \label{2ptfunOO}
\end{align}
We now derive the relation between $c_s$ and $c_f$. We start from the correlation function $\braket{\mathbb{O}^{a\dot a}(\tau)\bar{\mathbb{f}}^{\dot b b}(0)}_{\mathcal{W}}$ and we apply the supercharge $Q$. Using
\begin{equation}
 Q \mathbb{O}^{a\dot a}=\mathbb{f}^{a \dot a} \qquad  Q \bar {\mathbb{f}}^{\dot a a}=-2 \mathcal{D}_{\tau} \bar{\mathbb{O}}^{\dot a a}
\end{equation}
we get
\begin{equation}
 \braket{{\mathbb{f}}^{a\dot a}(\tau) \bar{\mathbb{f}}^{\dot b b}(0)}_{\mathcal{W}}-2 \braket{\mathbb{O}^{a\dot a}(\tau) \mathcal{D}_{\tau} \bar{\mathbb{O}}^{\dot b b}(0)}_{\mathcal{W}}=0
\end{equation}
which is equivalent to
\begin{equation}
 \braket{{\mathbb{f}}^{a\dot a}(\tau) \bar{\mathbb{f}}^{\dot b b}(0)}_{\mathcal{W}}= -2\pa_{\tau}\braket{\mathbb{O}^{a\dot a}(\tau)  \bar{\mathbb{O}}^{\dot b b}(0)}_{\mathcal{W}}
\end{equation}
and yields
\begin{equation}\label{cfcs}
 c_f=4 c_s\,.
\end{equation}

A similar strategy can be applied to the displacement supermultiplet. The correlation functions of its components are given by\footnote{The factor of 2 in the displacement two-point function is inserted to make contact with the standard definition of $C_D$ in the literature. Remember that $\mathbb{D}=\mathbb{D}_2-i  \mathbb{D}_3$ and $\braket{\mathbb{D} \bar{\mathbb{D}}}_{\mathcal{W}}=\braket{\mathbb{D}_2\mathbb{D}_2}_{\mathcal{W}}+\braket{\mathbb{D}_3\mathbb{D}_3}_{\mathcal{W}}=\frac{2 C_D}{|\tau|^4}$}
\begin{align}
 \braket{\mathbb{D}(\tau)\bar{\mathbb{D}}(0)}_{\mathcal{W}}&=\frac{2 C_D}{|\tau|^4} &
 \braket{\mathbb{F}(\tau)\bar{\mathbb{F}}(0)}_{\mathcal{W}}&=\frac{C_F \tau}{|\tau|^4}\,.
\end{align}
Starting from $\braket{\mathbb{F}(\tau)\bar{\mathbb{D}}(0)}_{\mathcal{W}}$ and using
\begin{equation}
 Q \mathbb{F}=2 \mathbb{D}  \qquad  Q \bar{\mathbb{D}}=-\mathcal{D}_{\tau} \bar{\mathbb{F} }
\end{equation}
 we find
\begin{equation}\label{CDCF}
 C_F=-\frac43 C_D\,.
\end{equation}
We determined a relation between the Zamolodchikov norm of operators in the same multiplet, which is equivalent to say that the superspace two-point function is fully determined up to an overall constant. In the next section we will see this explicitly.

\subsection{Correlation functions in superspace}\label{superspace}
The preserved superalgebra $su(1,1|1)$ has been widely studied in the context of $\mathcal{N}=2$ supersymmetric theory in two dimensions, where it appears as the global part of the $\mathcal{N}=2$ superconformal algebra in two dimensions when restricted to the holomorphic part. We can then use the results of \cite{Blumenhagen:1992sa,DiVecchia:1985ief,Mussardo:1988av} to write down defect correlation functions in superspace. We introduce the superspace coordinates $(\tau,\theta,\bar{\theta})$, where $\tau$ is a coordinate along the Wilson line while $\theta$ and $\bar \theta$ are Grassmann variables. A generic superfield reads
\begin{equation}\label{longsuperfield}
 \Phi_{\D,j}(\tau,\theta, \bar \theta)=\phi(\tau)+\theta \bar \chi(\tau) +\bar \theta \chi(\tau)+\theta \bar \theta \s(\tau)
\end{equation}
and the $su(1,1|1)$ generators act like differential operators (remember that for the representation theory of inserted operators the covariant derivative $\mathcal{D}_\tau$ has the role of the ordinary partial derivative, see \eqref{defprop})
\begin{align}
 P&=-\mathcal{D}_{\tau}&
  K&=-\tau^2 \mathcal{D}_{\tau}-\tau \theta \pa_{\theta}-\tau \bar \theta \pa_{\bar \theta}-2\tau \D-2 j \theta \bar \theta \nonumber\\
 D&=-\tau \mathcal{D}_{\tau}-\tfrac12 \theta \pa_\theta -\tfrac12 \bar \theta \pa_{\bar \theta}-\D&
 J&=\tfrac12 \theta \pa_{\theta}-\tfrac12 \bar \theta \pa_{\bar \theta}-j\nonumber\\
 Q&= \pa_{\bar \theta}-\theta \mathcal{D}_{\tau}&
 \bar Q&= \pa_{\theta}-\bar \theta \mathcal{D}_{\tau}\nonumber\\
 S&= \tau \pa_{\bar \theta}-\tau \theta \mathcal{D}_{\tau}-2 (\D+Q) \theta- \theta \bar \theta \pa_{\bar \theta}&
 \bar S&= \tau \pa_{\theta}-\tau \bar \theta \mathcal{D}_{\tau}-2 (\D-Q) \bar \theta+ \theta \bar \theta \pa_{\bar \theta} \,.\label{gendiff}
\end{align}
By applying these generators to \eqref{longsuperfield} one can easily check they respect the commutation relations \eqref{comm1}--\eqref{comm6}. The superspace is also equipped with supercovariant derivatives
\begin{align}\label{covder}
 \mathscr{D}&=\pa_{\theta}+\bar \theta \mathcal{D}_{\tau} &  \bar{\mathscr{D}}&=\pa_{\bar \theta}+ \theta \mathcal{D}_{\tau}\,. \qquad \qquad  
\end{align}
In this context short multiplets are represented as (anti)chiral superfields
\begin{align}
 &\mathcal{B}_{j}  & &\bar{\mathcal{B}}_{j}\\
 \bar{\mathscr{D}} \hat{\Phi}_{j,j}&=0 &  \mathscr{D} \hat{\Phi}_{-j,j}&=0
\end{align}
Chiral superfields $\hat \Phi_{j,j}$ depend only on the chiral coordinate $x=\tau+\theta\bar \theta$ (notice that $\bar{\mathscr{D}}x=0$) and on $\theta$. On the other hand, antichiral fields $\hat{\Phi}_{-j,j}$ depend only on $\tilde x=\tau-\theta\bar \theta$ and on $\bar \theta$. For the R-multiplet analyzed in section \eqref{scalarmult} we have the superfield expansions
\begin{align}
 \hat{\Phi}^{a\dot a}_{1,1}(x, \theta)&=\bar{\mathbb{O}}^{a \dot a}(x)+ \theta \bar{\mathbb{f}}^{a \dot a}(x)= \bar{\mathbb{O}}^{a \dot a}(\tau)+ \theta \bar{\mathbb{f}}^{a \dot a}(\tau)+\theta \bar \theta \mathcal{D}_1 \bar{\mathbb{O}}^{a \dot a}(\tau)\\
  \hat{\Phi}^{a\dot a}_{1,-1}(\tilde x, \bar{\theta})&=\mathbb{O}^{a \dot a}(\tilde x)+ \bar \theta \mathbb{f}^{a \dot a}(\tilde x)=\mathbb{O}^{a \dot a}(\tau)+ \bar \theta \mathbb{f}^{a \dot a}(\tau)-\theta \bar \theta \mathcal{D}_1 \mathbb{O}^{a \dot a}(\tau)\,.
\end{align}
For the displacement multiplet we have an additional factor of $2$ due to the transformation \eqref{QFD}
\begin{align}
  \hat{\Phi}_{\frac32,\frac32}(x, \theta)&=\bar{\mathbb{F}}(x)+2 \theta \bar{\mathbb{D}}(x)= \bar{\mathbb{F}}(\tau)+2 \theta \bar{\mathbb{D}}(\tau)+\theta \bar \theta \mathcal{D}_1 \bar{\mathbb{F}}(\tau)\\
  \hat{\Phi}_{\frac32,-\frac32}(\tilde x, \bar{\theta})&=\mathbb{F}(\tilde x)+2 \bar \theta \mathbb{D}(\tilde x)=\mathbb{F}(\tau)+ 2\bar \theta \mathbb{D}(\tau)-\theta \bar \theta \mathcal{D}_1 \mathbb{F}(\tau)\,.
\end{align}

As customary in the context of superconformal field theories, the structure of $n$-point correlation functions $\braket{\Phi(\tau_1,\theta_1,\bar\theta_1)\dots\Phi(\tau_n,\theta_n,\bar\theta_n)}_{\mathcal{W}}$ can be constrained by imposing invariance under the action of the eight generators \eqref{gendiff}. For instance, given a generator $J_i$ acting on the the superfield $\Phi(\tau_i,\theta_i,\bar{\theta_i})$ one imposes
\begin{equation}\label{confward}
 \sum_{i=1}^n J_i\braket{\Phi(\tau_1,\theta_1,\bar\theta_1)\dots\Phi(\tau_n,\theta_n,\bar\theta_n)}_{\mathcal{W}}=0\,.
\end{equation}
This provides a set of differential equations, which, for $n<4$ allow to completely fix the kinematical structure of the correlation function. Starting from the four-point correlation function one can form a set of superconformal invariants such that the requirement  \eqref{confward} is satisfied for an arbitrary function of such variables. Nevertheless, in this paper we will be dealing only with two- and three-point functions and we will be interested in the solution of the Ward identities \eqref{confward} for $n=2$ and $n=3$. These were studied in \cite{DiVecchia:1985ief,Mussardo:1988av}, but a complete solution was given only in \cite{Blumenhagen:1992sa}. 

The two-point function is non-zero only when the two superfields have opposite $U(1)$ charges and same conformal dimension
\begin{equation}\label{2pt}
 \braket{\Phi_{\D,j}(\tau_1,\theta_1,\bar \theta_1)\Phi_{\D,-j}(\tau_2,\theta_2,\bar \theta_2)}_{\mathcal{W}}=\frac{C_{\D,j}}{X_{12}^{2\D}} \left(1-2j\frac{\theta_{12} \bar\theta_{12}}{X_{12}}\right)
\end{equation}
where we introduced the variable $\theta_{ij}=\theta_i-\theta_j$ and the invariant distance 
\begin{equation}
 X_{ij}=\tau_i-\tau_j-\theta_i \bar \theta_j -\bar \theta_i \theta_j\,.
\end{equation}
One important feature of the two-point function \eqref{2pt} is that, for $\D=j$
\begin{equation}
\bar{\mathscr{D}}_1 \braket{\hat{\Phi}_{j,j}(\tau_1,\theta_1,\bar \theta_1)\hat{\Phi}_{j,-j}(\tau_2,\theta_2,\bar \theta_2)}_{\mathcal{W}}=\mathscr{D}_2 \braket{\hat{\Phi}_{j,j}(\tau_1,\theta_1,\bar \theta_1)\hat{\Phi}_{j,-j}(\tau_2,\theta_2,\bar \theta_2)}_{\mathcal{W}}=0
\end{equation}
which means that the expression on the r.h.s. of equation \eqref{2pt} applies also to the case of a chiral-antichiral two-point function. Therefore we can use it to rederive the relations of section \ref{Wardpres}. In particular
\begin{equation}
 \braket{\hat{\Phi}^{a\dot a}_{1,1} \hat {\Phi}^{b \dot b}_{1,-1}}_{\mathcal{W}}=\e^{ab} \e^{\dot a \dot b} C_{1,1} \left(\frac{1}{\t_{12}^2}+\frac{4 \theta_1 \bar{\theta}_2 }{\t_{12}^3}-\frac{2(\theta_1 \bar \theta_1+\theta_2 \bar \theta_2)}{\t_{12}^3}-\frac{6 \theta_1 \theta_2 \bar\theta_1 \bar \theta_2}{\t_{12}^4}\right)\,.
\end{equation}
Expanding the l.h.s. of this equation one finds
\begin{align}
 c_s&=C_{1,1}  & c_f=4 C_{1,1}
\end{align}
confirming equation \eqref{cfcs}. In a similar way
\begin{align}
 \braket{\hat{\Phi}_{\frac32,\frac32} \hat {\Phi}_{\frac32,-\frac32}}_{\mathcal{W}}=C_{\frac32,\frac32} \left(\frac{1}{\t_{12}^3}+\frac{6 \theta_1 \bar{\theta}_2 }{\t_{12}^4}-\frac{3(\theta_1 \bar \theta_1+\theta_2 \bar \theta_2)}{\t_{12}^4}-\frac{12 \theta_1 \theta_2 \bar\theta_1 \bar \theta_2}{\t_{12}^5}\right)
\end{align}
leading to
\begin{align}
 C_F&=-C_{\frac32,\frac32}  & C_D=\frac34 C_{\frac32,\frac32}
\end{align}
in agreement with \eqref{CDCF}.

The three-point function of long supermultiplets is non-vanishing for three different cases
\begin{align}
 j_1+j_2+j_3&=0  &  &\braket{\Phi_{\D_1,j_1}\Phi_{\D_2,j_2}\Phi_{\D_3,j_3}}_{\mathcal{W}}=c_{123} \frac{\left(1-2j_2\frac{\theta_{23}\bar \theta_{23}}{X_{23}}\right)  \left(1-2j_1\frac{\theta_{13}\bar \theta_{13}}{X_{13}}\right) }{X_{12}^{\D_{12}}X_{23}^{\D_{23}}X_{13}^{\D_{13}}}\ \label{tpf1} \\
 &&& \!\!\!+k_{123} \D_{12}\left[\frac{\left(\frac{\theta_{23}}{X_{23}}-\frac{\theta_{13}}{X_{13}}\right)\left(\frac{\bar\theta_{23}}{X_{23}}-\frac{\bar\theta_{13}}{X_{13}}\right)}{X_{12}^{\D_{12}+1}X_{23}^{\D_{23}-1}X_{13}^{\D_{13}-1}}-2\frac{\theta_{23}\bar\theta_{23} \theta_{13} \bar \theta_{13} \left(\frac{j_1}{X_{23}}+\frac{j_2}{X_{13}}\right)}{X_{12}^{\D_{12}+1}X_{23}^{\D_{23}}X_{13}^{\D_{13}}}\right]\nonumber \\
 j_1+j_2+j_3&=-\frac12  &  &\braket{\Phi_{\D_1,j_1}\Phi_{\D_2,j_2}\Phi_{\D_3,j_3}}_{\mathcal{W}}=c_{123}\frac{\frac{\bar \theta_{23}}{X_{23}}-\frac{\bar \theta_{13}}{X_{13}}+2\frac{j_2 \bar \theta_{13} \theta_{23} \bar \theta_{23}-j_1 \bar \theta_{23} \theta_{13} \bar \theta_{13}}{X_{23} X_{13}}}{X_{12}^{\D_{12}+\frac12}X_{23}^{\D_{23}-\frac12}X_{13}^{\D_{13}-\frac12}}\label{tpf2} \\
 j_1+j_2+j_3&=\frac12  &  &\braket{\Phi_{\D_1,j_1}\Phi_{\D_2,j_2}\Phi_{\D_3,j_3}}_{\mathcal{W}}=c_{123}\frac{\frac{ \theta_{23}}{X_{23}}-\frac{ \theta_{13}}{X_{13}}+2\frac{j_2  \theta_{13} \theta_{23} \bar \theta_{23}-j_1  \theta_{23} \theta_{13} \bar \theta_{13}}{X_{23} X_{13}}}{X_{12}^{\D_{12}+\frac12}X_{23}^{\D_{23}-\frac12}X_{13}^{\D_{13}-\frac12}}\label{tpf3} 
\end{align}
where $\D_{12}=\D_1+\D_2-\D_3$, $\D_{13}=\D_1+\D_3-\D_2$, $\D_{23}=\D_2+\D_3-\D_1$ while $c_{123}$ and $k_{123}$ are undetermined parameters. In the following we will be interested in three-point functions involving short multiplets, therefore we would like to understand which constraints need to be imposed on the expressions \eqref{tpf1}, \eqref{tpf2}, \eqref{tpf3} when chiral and antichiral superfields are involved. To do this we need to apply the covariant derivatives \eqref{covder} on the explicit expression of the three-point functions and check whether they are annihilated\footnote{We are grateful to Madalena Lemos for valuable help and useful suggestions on this point.}. For the case of vanishing total R-charge, which depends on two free parameters, this imposes constraints on the $c_{123}$ and $k_{123}$. The list of non-vanishing three-point functions is

{\footnotesize
\begin{center}
\begin{tabular}{|c|c|c|}
\hline
\multicolumn{3}{|c|}{$j_1+j_2+j_3=0$}\\
\hline
 Condition & Multiplet type & Three-point function \\
 \hline
 $k_{123}=-c_{123}$ & $\braket{\mathcal{B}_{j_1} \mathcal{A}_{\D_2, j_2} \mathcal{A}_{\D_3, j_3}}$ & $\braket{\hat{\Phi}_{j_1,j_1}\Phi_{\D_2,j_2}\Phi_{\D_3,j_3}}_{\mathcal{W}}$\\
  $k_{123}=c_{123}$ & $\braket{\bar{\mathcal{B}}_{j_1} \mathcal{A}_{\D_2, j_2} \mathcal{A}_{\D_3, j_3}}$ & $\braket{\hat{\Phi}_{-j_1,j_1}\Phi_{\D_2,j_2}\Phi_{\D_3,j_3}}_{\mathcal{W}}$\\
   $k_{123}=c_{123}$ & $\braket{\mathcal{A}_{\D_1, j_1} \mathcal{B}_{j_2} \mathcal{A}_{\D_3, j_3}}$ & $\braket{\Phi_{\D_1,j_1}\hat{\Phi}_{j_2,j_2}\Phi_{\D_3,j_3}}_{\mathcal{W}} $\\
   $k_{123}=-c_{123}$ & $\braket{\mathcal{A}_{\D_1, j_1} \bar{\mathcal{B}}_{j_2} \mathcal{A}_{\D_3, j_3}}$ & $\braket{\Phi_{\D_1,j_1}\hat{\Phi}_{-j_2,j_2}\Phi_{\D_3,j_3}}_{\mathcal{W}} $\\
   $k_{123}=\frac{\D_2-\D_1+j_2-j_1}{\D_2+\D_1+j_2+j_1}c_{123}$ & $\braket{\mathcal{A}_{\D_1, j_1} \mathcal{A}_{\D_2, j_2}\mathcal{B}_{j_3} }$ & $\braket{\Phi_{\D_1,j_1}\Phi_{\D_2,j_2}\hat{\Phi}_{j_3,j_3}}_{\mathcal{W}} $\\
   $k_{123}=-\frac{\D_2-\D_1-j_2+j_1}{\D_2+\D_1-j_2-j_1}c_{123}$ & $\braket{\mathcal{A}_{\D_1, j_1} \mathcal{A}_{\D_2, j_2} \bar{\mathcal{B}}_{j_3} }$ & $\braket{\Phi_{\D_1,j_1}\Phi_{\D_2,j_2}\Phi_{-j_3,j_3}}_{\mathcal{W}} $\\
   $k_{123}=-c_{123}$ & $\braket{\mathcal{B}_{j_1} \bar{\mathcal{B}}_{j_2} \mathcal{A}_{\D_3, j_3}}$ & $\braket{\hat{\Phi}_{j_1,j_1}\hat{\Phi}_{-j_2,j_2}\Phi_{\D_3,j_3}}_{\mathcal{W}}$\\
   $k_{123}=c_{123}$ & $\braket{\bar{\mathcal{B}}_{j_1} \mathcal{B}_{j_2} \mathcal{A}_{\D_3, j_3}}$ & $\braket{\hat{\Phi}_{-j_1,j_1}\hat{\Phi}_{j_2,j_2}\Phi_{\D_3,j_3}}_{\mathcal{W}}$\\
   $k_{123}=-c_{123}$ & $\braket{\mathcal{B}_{j_1} \mathcal{A}_{\D_2, j_2} \bar{\mathcal{B}}_{j_3}}$ & $\braket{\hat{\Phi}_{j_1,j_1}\Phi_{\D_2,j_2}\hat{\Phi}_{-j_3,j_3}}_{\mathcal{W}}$\\
   $k_{123}=c_{123}$ & $\braket{\bar{\mathcal{B}}_{j_1} \mathcal{A}_{\D_2, j_2} \mathcal{B}_{j_3}}$ & $\braket{\hat{\Phi}_{-j_1,j_1}\Phi_{\D_2,j_2}\hat{\Phi}_{j_3,j_3}}_{\mathcal{W}}$\\
    $k_{123}=c_{123}$ & $\braket{\mathcal{A}_{\D_1, j_1} \mathcal{B}_{j_2} \bar{\mathcal{B}}_{j_3}}$ & $\braket{\Phi_{\D_1,j_1}\hat{\Phi}_{j_2,j_2}\hat{\Phi}_{-j_3,j_3}}_{\mathcal{W}}$\\
     $k_{123}=-c_{123}$ & $\braket{\mathcal{A}_{\D_1, j_1} \bar{\mathcal{B}}_{j_2} \mathcal{B}_{j_3}}$ & $\braket{\Phi_{\D_1,j_1}\hat{\Phi}_{-j_2,j_2}\hat{\Phi}_{j_3,j_3}}_{\mathcal{W}}$\\
   $k_{123}=c_{123}$ & $\braket{\bar{\mathcal{B}}_{j_1} \mathcal{B}_{j_2} \mathcal{B}_{j_3}}$ & $\braket{\hat{\Phi}_{-j_1,j_1}\hat{\Phi}_{j_2,j_2}\hat{\Phi}_{j_3,j_3}}_{\mathcal{W}}$\\
   $k_{123}=-c_{123}$ & $\braket{\mathcal{B}_{j_1} \bar{\mathcal{B}}_{j_2} \bar{\mathcal{B}}_{j_3}}$ & $\braket{\hat{\Phi}_{j_1,j_1}\hat{\Phi}_{-j_2,j_2}\hat{\Phi}_{-j_3,j_3}}_{\mathcal{W}}$\\
   $k_{123}=c_{123}$ & $\braket{\bar{\mathcal{B}}_{j_1} \mathcal{B}_{j_2} \bar{\mathcal{B}}_{j_3}}$ & $\braket{\hat{\Phi}_{-j_1,j_1}\hat{\Phi}_{j_2,j_2}\hat{\Phi}_{-j_3,j_3}}_{\mathcal{W}}$\\
    $k_{123}=-c_{123}$ & $\braket{\mathcal{B}_{j_1} \bar{\mathcal{B}}_{j_2} \mathcal{B}_{j_3}}$ & $\braket{\hat{\Phi}_{j_1,j_1}\hat{\Phi}_{-j_2,j_2}\hat{\Phi}_{j_3,j_3}}_{\mathcal{W}}$\\
  \hline
\end{tabular}

\end{center}
}

On the other hand, when the R-charges do not sum to zero there is only one free parameter, therefore the three-point function either vanishes or is given by equations \eqref{tpf2} and \eqref{tpf3}. The list of non-vanishing three-point functions in this case is
{\footnotesize
\begin{center}
\begin{tabular}{|c|c|}
\hline
\multicolumn{2}{|c|}{$j_1+j_2+j_3=-\frac12$}\\
\hline
 Multiplet type & Three-point function \\
 \hline
   $\braket{\bar{\mathcal{B}}_{j_1} \mathcal{A}_{\D_2, j_2} \mathcal{A}_{\D_3, j_3}}$ & $\braket{\hat{\Phi}_{-j_1,j_1}\Phi_{\D_2,j_2}\Phi_{\D_3,j_3}}_{\mathcal{W}}$\\
   $\braket{\mathcal{A}_{\D_1, j_1} \bar{\mathcal{B}}_{j_2} \mathcal{A}_{\D_3, j_3}}$ & $\braket{\Phi_{\D_1,j_1}\hat{\Phi}_{-j_2,j_2}\Phi_{\D_3,j_3}}_{\mathcal{W}} $\\
   $\braket{\mathcal{A}_{\D_1, j_1} \mathcal{A}_{\D_2, j_2} \bar{\mathcal{B}}_{j_3} }$ & $\braket{\Phi_{\D_1,j_1}\Phi_{\D_2,j_2}\Phi_{-j_3,j_3}}_{\mathcal{W}} $\\
  $\braket{\bar{\mathcal{B}}_{j_1} \bar{\mathcal{B}}_{j_2} \mathcal{A}_{\D_3, j_3}}$ & $\braket{\hat{\Phi}_{-j_1,j_1}\hat{\Phi}_{-j_2,j_2}\Phi_{\D_3,j_3}}_{\mathcal{W}}$\\
    $\braket{\bar{\mathcal{B}}_{j_1} \mathcal{A}_{\D_2, j_2} \bar{\mathcal{B}}_{j_3}}$ & $\braket{\hat{\Phi}_{-j_1,j_1}\Phi_{\D_2,j_2}\hat{\Phi}_{-j_3,j_3}}_{\mathcal{W}}$\\
     $\braket{\mathcal{A}_{\D_1, j_1} \bar{\mathcal{B}}_{j_2} \bar{\mathcal{B}}_{j_3}}$ & $\braket{\Phi_{\D_1,j_1}\hat{\Phi}_{-j_2,j_2}\hat{\Phi}_{-j_3,j_3}}_{\mathcal{W}}$\\
  \hline
\end{tabular}
\begin{tabular}{|c|c|}
\hline
\multicolumn{2}{|c|}{$j_1+j_2+j_3=\frac12$}\\
\hline
 Multiplet type & Three-point function \\
 \hline
   $\braket{\mathcal{B}_{j_1} \mathcal{A}_{\D_2, j_2} \mathcal{A}_{\D_3, j_3}}$ & $\braket{\hat{\Phi}_{j_1,j_1}\Phi_{\D_2,j_2}\Phi_{\D_3,j_3}}_{\mathcal{W}}$\\
   $\braket{\mathcal{A}_{\D_1, j_1} \mathcal{B}_{j_2} \mathcal{A}_{\D_3, j_3}}$ & $\braket{\Phi_{\D_1,j_1}\hat{\Phi}_{j_2,j_2}\Phi_{\D_3,j_3}}_{\mathcal{W}} $\\
   $\braket{\mathcal{A}_{\D_1, j_1} \mathcal{A}_{\D_2, j_2} \mathcal{B}_{j_3} }$ & $\braket{\Phi_{\D_1,j_1}\Phi_{\D_2,j_2}\Phi_{j_3,j_3}}_{\mathcal{W}} $\\
  $\braket{\mathcal{B}_{j_1} \mathcal{B}_{j_2} \mathcal{A}_{\D_3, j_3}}$ & $\braket{\hat{\Phi}_{j_1,j_1}\hat{\Phi}_{j_2,j_2}\Phi_{\D_3,j_3}}_{\mathcal{W}}$\\
    $\braket{\mathcal{B}_{j_1} \mathcal{A}_{\D_2, j_2} \mathcal{B}_{j_3}}$ & $\braket{\hat{\Phi}_{j_1,j_1}\Phi_{\D_2,j_2}\hat{\Phi}_{j_3,j_3}}_{\mathcal{W}}$\\
     $\braket{\mathcal{A}_{\D_1, j_1} \mathcal{B}_{j_2} \mathcal{B}_{j_3}}$ & $\braket{\Phi_{\D_1,j_1}\hat{\Phi}_{j_2,j_2}\hat{\Phi}_{j_3,j_3}}_{\mathcal{W}}$\\
  \hline
\end{tabular}
\end{center}
}
One important consequence of these conditions is that no three-point function with non-vanishing total r-charge involves a chiral and an antichiral superfield, in particular 
\begin{equation}\label{crucial}
 \braket{\hat{\Phi}_{-j_1,j_1}\hat{\Phi}_{-j_2,j_2}\hat{\Phi}_{j_3,j_3}}_{\mathcal{W}}=0 \qquad \text{for} \qquad j_1+j_2+j_3=\pm \frac12\,.
\end{equation}
This feature of the $su(1,1|1)$ invariant three-point functions will be very useful in the following.

\subsection{Ward identity with broken supercharges}\label{Wardbrok}
We now consider the action of a broken supercharge on a defect correlation function in order to find a simple relation between $c_s$ and $C_D$, i.e. between the two Bremsstrahlung functions $B_{\theta}$ and $B_{\phi}$. Let us stress this is a very peculiar feature of the defect setup, since in the original theory, unless it originates from the breaking of some larger supersymmetry, it wouldn't be possible to connect objects in different supermultiplets. We first observe that the supercharge $\mathfrak{Q}^{a \dot a}$ connects the two supermultiplets since    
\begin{equation}
  \e_{ab} \e_{\dot a \dot b}\mathfrak{Q}^{a \dot a} \mathbb{O}^{b\dot b}=2\, \mathbb{F}\,.
\end{equation}
This means that in the original theory the two operators belonged to the same supermultiplet.
\begin{center}
\begin{tikzpicture}
  \node[above] at (0,4.8) {$\mathbb{O}^{a\dot a}$};
  \draw[->]  (0.35,4.8)--(0.65,4.5);
  \draw[->,dotted]  (-0.35,4.8)--(-0.65,4.5);
   \draw[->,dotted]  (0.65,3.8)--(0.35,3.5);
  \draw[->]  (-0.65,3.8)--(-0.35,3.5);
  \node[above] at (1,3.8) {$\mathbb{f}^{a \dot a}$};
  \node[above] at (-1,3.8) {$\mathbb{F}$};
  \node[above] at (0,2.8) {$\mathbb{D}$};
    \node at (0,2) {$-1$};
  \node at (1,2) {$-\frac12$};
   \node at (-1,2) {$-\frac32$};
  \node at (-2,2.5) {$\D$};
  \node at (-1.6,2) {$j$};
  \draw[-] (-1.6,2.4)--(-2,2);
   \node at (-2,4.1) {$\frac32$};
   \node at (-2,5.1) {$1$};
   \node at (-2,3.1) {$2$};
    \node at (0.7,4.9) {$\scriptstyle{Q}$};
  \node at (-0.7,4.9) {$\scriptstyle{\mathfrak{Q}^{a\dot a}}$};
  \end{tikzpicture}\hspace{1.6 cm}
\end{center}
When the supercharge is broken one cannot derive a Ward identity like we did in section \ref{Wardpres}, where we tacitly assumed that $Q$ annihilates the Wilson line, but one has to rely on the relation \eqref{vivalatopa} for a supersymmetry variation. For the two-point function $\braket{\mathbb{O}^{a \dot a}(\t_1) \bar{\mathbb{F}}(\tau_2)}_{\mathcal{W}}$ we have
\begin{multline}
  \braket{\mathfrak{Q}_{a \dot a}\mathbb{O}^{a \dot a}(\tau_1) \bar{\mathbb{F}}(\tau_2)}_{\mathcal{W}} +  \braket{\mathbb{O}^{a \dot a}(\tau_1) \mathfrak{Q}_{a \dot a}\bar{\mathbb{F}}(\tau_2)}_{\mathcal{W}}=i \int_{-\infty}^{\tau_1} d\tau \braket{\mathbb{f}_{a \dot a}(\tau)\mathbb{O}^{a \dot a}(\tau_1)\bar{\mathbb{F}}(\tau_2)}_{\mathcal{W}}\\+i\int_{\tau_1}^{\tau_2} d\tau \braket{\mathbb{O}^{a \dot a}(\tau_1)\mathbb{f}_{a \dot a}(\tau)\bar{\mathbb{F}}(\tau_2)}_{\mathcal{W}}-i\int_{\tau_2}^{\infty} d\tau \braket{\mathbb{O}^{a \dot a}(\tau_1)\bar{\mathbb{F}}(\tau_2)\mathbb{f}_{a \dot a}(\tau)}_{\mathcal{W}}\,.
\end{multline}
A quick look at the three-point functions show that they appear in the component expansion of the superfield correlation function
\begin{align}
 \braket{\hat\Phi_{1,-1}^{a \dot a}(\tau_1,\theta_1,\bar \theta_1) \hat\Phi_{1,-1}^{b \dot b}(\tau, \theta, \bar \theta) \hat\Phi_{\frac32,\frac32}(\tau_2,\theta_2,\bar \theta_2)}_{\mathcal{W}}\sim \bar\theta \braket{\mathbb{O}^{a \dot a}(\tau_1)\mathbb{f}^{b \dot b}(\tau)\bar{\mathbb{F}}(\tau_2)}_{\mathcal{W}} + \dots
\end{align}
which vanishes according to \eqref{crucial}. Therefore we are left with
\begin{align}
 \braket{\mathfrak{Q}_{a \dot a}\mathbb{O}^{a \dot a}(\tau_1) \bar{\mathbb{F}}(\tau_2)}_{\mathcal{W}}=-  \braket{\mathbb{O}^{a \dot a}(\tau_1) \mathfrak{Q}_{a \dot a}\bar{\mathbb{F}}(\tau_2)}_{\mathcal{W}}\,.
\end{align}
Using the transformations in appendix \ref{broken}, one finds
\begin{align}
 & \mathfrak{Q}^{a \dot a} \bar{\mathbb{F}}=-2 \mathcal{D}_{\tau} \bar{\mathbb{O}}^{\dot a a}-\frac{4 \pi i}{k}\mathcal{C}^{a \dot a}
\end{align}
where $\mathcal{C}^{a \dot a}$ is a conformal primary (though not necessarily superconformal primary) of classical dimension $2$, whose explicit form is given in appendix \ref{broken} and it is not needed here since its two-point function with $\mathbb{O}^{a\dot a}$ is clearly vanishing. Therefore keeping only the derivative term we get
\begin{align}
 \braket{\mathbb{F}(\tau_1) \bar {\mathbb{F}}(\tau_2)}_{\mathcal{W}}=\e_{ab} \e_{\dot a \dot b} \pa_{\tau_2} \braket{ \mathbb{O}^{a \dot a}(\tau_1)\mathbb{O}^{b \dot b}(\tau_2)}_{\mathcal{W}}
\end{align}
and consequently
\begin{align}\label{CDcs}
 C_F= 8 c_s \quad \Rightarrow \quad C_D=-6 c_s\,.
\end{align}
In order to relate this result with the Bremsstrahlung functions we need to use the universal result of \cite{Correa:2012at}
\begin{align}
 C_D=12 B_{1/6}^{\varphi} 
\end{align}
and the relation found in \cite{Aguilera-Damia:2014bqa}\footnote{Notice that the operators $\mathbb{O}^{a \dot a}$ in \cite{Aguilera-Damia:2014bqa} are normalized differently (by a factor $\frac{4\pi i}{k}$) and therefore the parameter $\gamma$ in equation (62) and (65) of \cite{Aguilera-Damia:2014bqa} is $\g=\frac{-k^2}{16\pi^2} c_s$}
\begin{equation}
 c_s=-4 B_{1/6}^{\theta}\,.
\end{equation}
Inserting them inside \eqref{CDcs} we get the main result of this paper
\begin{align}\label{Bbosonicrel}
 B_{1/6}^{\varphi}=2B_{1/6}^{\theta}\,.
\end{align}

\section{Relations between Bremsstrahlung functions}\label{sec:5}
In this section we review the cohomological relation between bosonic and fermionic supersymmetric Wilson loops, also introducing the framing regularization. Moreover, we explore some consequences of the result of the previous section leading to a relation between all the Bremsstrahlung functions.

\subsection{The cohomological equivalence and the framing}\label{sec:framing}

Since the supercharges preserved by the bosonic Wilson loop are shared by the fermionic one, one may ask whether there is a relation between these operators. This possibility was explored in \cite{Drukker:2009hy}, where it was shown that they are in the same cohomology class under the shared supercharges. This means that the difference between the two Wilson loops is exact with respect to a linear combination $\tilde{Q}$ of the shared supercharges, namely there exists a functional $V$ of the fields such that
\begin{equation}\label{cohomo}
\mathcal{W}_F[\mathcal{C}]-\tilde{\mathcal{W}}_B[\mathcal{C}]=\tilde{Q}\,V
\end{equation}
where $\tilde{\mathcal{W}}_B$ is a linear combination of the operators $\mathcal{W}_B$ and $\hat{\mathcal{W}}_B$.

The simplest example is given by the most supersymmetric case, when the contour is an infinite straight line or a maximal circle. In this case the operators in \eqref{cohomo} are given by
\begin{equation}\label{opLC}
\mathcal{W}_F[\mathcal{C}]=\mathcal{W}_{1/2}\,,\qquad
\tilde{\mathcal{W}}_B[\mathcal{C}]=\frac12\left(\mathcal{W}_{1/6}+\hat{\mathcal{W}}_{1/6}\right)\,.
\end{equation} 
The combination of charges $\tilde{Q}$ is obviously different in the two cases: for the straight line $\tilde{Q}=Q+\bar{Q}$ where $Q$ and $\bar{Q}$ are defined in section \ref{symmetry}, for the maximal circle $\tilde{Q}$ is a combination of the Poincar\'e and superconformal supercharges (see \cite{Drukker:2009hy}).  
For less supersymmetric Wilson loops the relation \eqref{cohomo} needs to be modified. For instance, considering the 1/6 and 1/12 BPS latitudes, we have 
\begin{equation}\label{oplat}
\mathcal{W}_F[\mathcal{C}]=\mathcal{W}_{F}(\nu)\,,\qquad
\tilde{\mathcal{W}}_B[\mathcal{C}]=\frac{e^{-i\frac{\pi\nu}{2}}\mathcal{W}_{B}(\nu)-e^{i\frac{\pi\nu}{2}}\hat{\mathcal{W}}_{B}(\nu)}{e^{-i\frac{\pi\nu}{2}}-e^{i\frac{\pi\nu}{2}}}\,.
\end{equation} 
In this case the combination of charges $\tilde{Q}$ is given in \cite{Bianchi:2014laa}. To relate the vacuum expectation value of bosonic and fermionic Wilson operators using \eqref{cohomo}, we need to review the framing regularization.

As in pure Chern-Simons theory, in ABJM theory the expectation value of Wilson loops is affected by finite regularization ambiguities when short-distance divergences of gauge field correlators are treated with a point-splitting regularization. To avoid this, one requires that each point runs on a different path (frame). The new path can be written as
\begin{equation}
\mathcal{C}_m:\qquad x^\mu(\tau)\;\rightarrow\;y^\mu(\tau)=x^\mu(\tau)+\alpha \,m^\mu(\tau)\,,\qquad |m(\tau)|=1
\end{equation}
where $m^\mu(\tau)$ is orthogonal to the contour $\mathcal{C}$. The expectation value of the Wilson loop does not depend on the choice of the framing vector $m^\mu(\tau)$, but only on the cotorsion $f$, i.e. the number of times  the modified contour $\mathcal{C}_m$ winds around the original one $\mathcal{C}$. In pure Chern-Simons theory, the contribution of the framing is captured by a one-loop exact overall phase. 

In ABJM the effect of the framing on BPS Wilson loops is more subtle and it was studied in great detail in \cite{Bianchi:2014laa,Bianchi:2016yzj,Bianchi:2016gpg}. One can split the bosonic and fermionic Wilson loops into their phases and moduli
\begin{equation}\begin{split}\label{framingphase}
&\langle\mathcal{W}_B[\mathcal{C}]\rangle_f=e^{i\Phi_B(f,\mathcal{C})}\,|\langle\mathcal{W}_B[\mathcal{C}]\rangle|\,,\qquad
\langle\hat{\mathcal{W}}_B[\mathcal{C}]\rangle_f=e^{-i\Phi_B(f,\mathcal{C})}\,|\langle\hat{\mathcal{W}}_B[\mathcal{C}]\rangle|\,,\\
&\langle\mathcal{W}_F[\mathcal{C}]\rangle_f=\langle\mathcal{W}_F[\mathcal{C}]\rangle_0\,,
\end{split}\end{equation}
where $\langle...\rangle_f$ denotes the expectation value at framing $f$, whose dependence is completely encoded in the the phase $\Phi_B$. For this reason, the absolute value $|\langle\mathcal{W}_B[\mathcal{C}]\rangle |$ does not depend on $f$ and can be taken at any desired framing. Notice that this is no longer true when considering arbitrary winding $n$. Furthermore, compared to \cite{Bianchi:2016yzj}\footnote{Hence we do not refer to our $\Phi_B$ as ``framing phase'' to avoid confusion with previous literature. We thank the authors of \cite{Bianchi:2018bke} for private communication on this point.}, we do not identify the absolute value $|\langle\mathcal{W}_B[\mathcal{C}]\rangle|$ with the expectation value at framing 0, which we allow to be complex. 
We stress also, that the phase is a non-trivial function of $\lambda$, which we omit from its arguments for simplicity. For the fermionic Wilson loop, we take into account that in \cite{Drukker:2010nc} the phase, which arises naturally in the computation of the loop via localization, is argued to coincide with that computed in Chern-Simons theory with gauge supergroup $U(N|N)$, which in turn equals zero. As a result, $\langle\mathcal{W}_F[\mathcal{C}]\rangle_f$ is insensitive to $f$.

In \cite{Drukker:2009hy, Drukker:2010nc} it was observed that the quantum realization of the relation \eqref{cohomo} for the circular Wilson loops needs a specific choice of framing. For example the relation
\begin{equation}\label{rel1}
\langle\mathcal{W}_F(1)\rangle_1=\frac{1}{2}\left(\langle\mathcal{W}_B(1)\rangle_1+\langle\hat{\mathcal{W}}_B(1)\rangle_1\right)
\end{equation}
computed by means of supersymmetric localization \cite{Kapustin:2009kz} shows clearly this feature. 
In the latitude Wilson loop case a similar situation is expected \cite{Bianchi:2014laa}:
\begin{equation}\label{rel3}
\langle\mathcal{W}_F(\nu)\rangle_\nu=\frac{e^{-i\frac{\pi\nu}{2}}\langle\mathcal{W}_B(\nu)\rangle_\nu-e^{i\frac{\pi\nu}{2}}\langle\hat{\mathcal{W}}_B(\nu)\rangle_\nu}{e^{-i\frac{\pi\nu}{2}}-e^{i\frac{\pi\nu}{2}}}\,,
\end{equation}
given the relations \eqref{framingphase} evaluated at framing $f=\nu$ and the definitions \eqref{oplat}.

\subsection{Bremsstrahlung functions as bosonic phase}
In the following the main goal is to work out some consequences of the result \eqref{Bbosonicrel}.
Combining this with \eqref{B_sixth_varphi} and \eqref{B_sixth_theta_3} leads to
\begin{equation}\label{label4}
B^\theta_{1/6}=\frac{1}{4\pi^2} \Big.\pa_{\nu} \log |\braket{\mathcal{W}_{B}(\nu)}_0|\Big|_{\nu=1}=\frac{1}{8\pi^2} \pa_{n} \log |\braket{\mathcal{W}^{n}_{B}(1)}_1|\Big|_{n=1}=\frac12 B^\varphi_{1/6}\,.
\end{equation}
Next we adapt \eqref{label4} to the bosonic latitude
\begin{equation}\label{framingphase2}\begin{split}
&\langle\mathcal{W}_B(\nu)\rangle_f=e^{i\Phi_B(f,\nu)}\,|\langle\mathcal{W}_B(\nu)\rangle|
\end{split}\end{equation}
and write an equation involving latitude and $n$-wound circle 
\begin{equation}\label{label3.1}
|\braket{\mathcal{W}_{B}(\nu)}_0|=|\braket{\mathcal{W}^{n}_{B}(1)}_1|
\end{equation}
where the left-hand side can be taken at any framing, including $f=1$. The relation \eqref{label3.1} serves as implicit definition of the real function $n=n(\nu)$\footnote{\label{2014} It does not have to coincide with the function of the same name in \cite{Bianchi:2014laa}, which our \eqref{B_half_3} is quoted from, defined from the latitude at framing $f=\nu$ and no absolute values in \eqref{label3.1}.}.
Unfortunately, such function cannot be truly known unless the latitude loop is known exactly.
Nevertheless, from equation \eqref{label4} we can conclude that
\begin{equation}\label{data}
n(1)=1\,,\qquad \frac{\partial n(\nu)}{\partial \nu}\Big|_{\nu=1}=\frac12\,.
\end{equation}

 We now recall equation \eqref{B_half_1}, where $B_{1/2}$ is expressed as logarithmic derivative of the fermionic latitude respect to the latitude parameter $\nu$. Leaving aside the possibility of accessing the latter exactly,
the cohomological relation \eqref{rel3} allows to argue that the derivative is expressible as a combination of the bosonic operators, associated to the $U(N)_k$ and $U(N)_{-k}$ gauge group factors, on the same latitude contour  \cite{Bianchi:2014laa}:
\begin{equation}\label{B_half_2}
B_{1/2}=\frac{1}{4\pi^2}\left[\Big.\partial_\nu \log\left(\langle \mathcal{W}_B(\nu) \rangle_\nu+\langle\hat{\mathcal{W}}_B(\nu)\rangle_\nu \right)\Big|_{\nu=1} +\frac{\pi}{2}\tan\Phi_B \right]\,.
\end{equation}
The function $\Phi_B\equiv\Phi_B(1,1)$ from \eqref{framingphase2} is the phase of the expectation value of $\mathcal{W}_B(1)$ on the maximal circle of $S^2$ ($\nu=1$) as computed by localization ($f=1$) \cite{Kapustin:2009kz,Marino:2009jd,Drukker:2010nc}.  
Then it is easy to see that the second term in \eqref{B_half_2} can be written as
\begin{equation}\label{tanPhi}
\tan\Phi_B
=-i \frac{\langle\mathcal{W}_B(1)\rangle_1-\langle\hat{\mathcal{W}}_B(1)\rangle_1}{\langle\mathcal{W}_B(1)\rangle_1+\langle\hat{\mathcal{W}}_B(1)\rangle_1}
=-i \frac{\langle\mathcal{W}_B(1)\rangle_1-\langle\mathcal{W}_B(1)\rangle^*_1}{\langle\mathcal{W}_B(1)\rangle_1+\langle\mathcal{W}_B(1)\rangle^*_1}
\,,
\end{equation}
where we take into account $\braket{\hat{\mathcal{W}}_{B}(\nu)}_{\nu}= (\braket{\mathcal{W}_{B}(\nu)}_{\nu})^*$ and use the the shorthand $\langle \mathcal{W}_B(1)\rangle_1=\langle \mathcal{W}_B^1(1)\rangle_1$ for the 1-wound circle. On the contrary, the first term is unknown because localization computes the bosonic latitude only in the great circle case. However, we can repeat the steps in \cite{Bianchi:2014laa}: trade the latitude parameter with the winding number (albeit in a way different from \eqref{label3.1}, see footnote \ref{2014} again)
\begin{equation}\label{B_half_3}
\Big[\partial_\nu \log\left(\langle\mathcal{W}_B(\nu)\rangle_\nu+\langle\hat{\mathcal{W}}_B(\nu)\rangle_\nu \right)\Big]_{\nu=1}
\!\!\!\!=
\Big.\left[\partial_n \log\left(\langle \mathcal{W}_B^{n}(1)\rangle_1+\langle\hat{\mathcal{W}}_B^{n}(1)\rangle_1 \right) \frac{\partial n(\nu)}{\partial \nu}\right]_{n=1}
\end{equation}
and use the localization formula for the $n$-wound loop to check that the logarithmic derivative in the right-hand side of \eqref{B_half_3} vanishes\footnote{More generally, the vanishing for any $\nu$ is a consequence of the conjectured matrix model of \cite{Bianchi:2018bke}.}. Although this argument does not require an explicit $n(\nu)$, one may worry that a divergent value of ${\partial n(\nu)}/{\partial \nu}|_{n=1}$ prevents the left-hand side of \eqref{B_half_3} from being zero as well. While this possibility was somewhat excluded in \cite{Bianchi:2014laa}, we may use the behavior of its analogue function given in \eqref{data} and arrive to
\begin{equation}\label{B_half_4}
B_{1/2}=\frac{1}{8\pi}\tan \Phi_B
=-\frac{i}{8 \pi} \frac{\langle\mathcal{W}_B(1)\rangle_1-\langle\mathcal{W}_B(1)\rangle^*_1}{\langle\mathcal{W}_B(1)\rangle_1+\langle\mathcal{W}_B(1)\rangle^*_1}\,.
\end{equation}

The weak-coupling expansion of the $n$-wound loop in the planar limit reveals that \eqref{B_half_3} should be described by an even function of $\lambda$ whereas $\langle\mathcal{W}_B(1)\rangle_1$ contains odd powers only. Thus the vanishing of \eqref{B_half_3} leads to a Bremsstrahlung function that is an odd real function of the coupling. The final expression \eqref{B_half_4} agrees with the early prediction in \cite{Lewkowycz:2013laa} motivated by different arguments and it was thoroughly checked with perturbative data to three loops at weak coupling for finite $N$ \cite{Griguolo:2012iq,Bianchi:2014laa,Bianchi:2017svd} and at classical and one-loop order at large coupling \cite{Forini:2012bb,Correa:2014aga,Aguilera-Damia:2014bqa}\footnote{An early mismatch with the analysis of \cite{Forini:2012bb} was resolved by a careful re-evaluation of the one-loop determinants for string fluctuations \cite{Aguilera-Damia:2014bqa}.}. In section \ref{matrix_model} we derive an exact closed-form for \eqref{B_half_4} using the results of localization.

The vanishing of the left-hand side of \eqref{B_half_3} has another interesting implication. Using \eqref{framingphase2} it becomes
\begin{equation}
\label{label7}
\begin{split}
\Big[\pa_{\nu}  \log |\braket{\mathcal{W}_{B}(\nu)}_0 | \Big]_{\nu=1}=\tan\Phi_B \,\Big[ \partial_\nu \Phi_B(\nu,\nu) \Big]_{\nu=1}\,.
\end{split}
\end{equation}
with the left-hand side taken at framing zero since, as mentioned before, the absolute value is framing independent.
The combination of the proven relation \eqref{Bbosonicrel} and \eqref{B_half_4} leads to 
\begin{equation}
\label{label8}
\begin{split}
B^\theta_{1/6}=\frac{1}{2}B^\varphi_{1/6}
=\frac{2}{\pi} B_{1/2}\, \partial_\nu \Phi_B(\nu,\nu)|_{\nu=1}\,.
\end{split}
\end{equation}
This chain of equalities is the main result of the section: the Bremsstrahlung functions in ABJM are all related. 
In particular, they can all be expressed all in terms of an unique function $\Phi_B(\nu,\nu)$, the phase defined in \eqref{framingphase2}. We remark that the statement \eqref{label8} is highly non-trivial because it establishes a connection between different  small-angle deformations of Wilson lines preserving different degrees of supersymmetry. 

Given \eqref{label8}, we predict the first few orders of $\Phi_B(\nu,\nu)$ at weak coupling and large $N$\footnote{We are thankful to Luca Griguolo and Domenico Seminara for discussions on this point.}. The bosonic latitude at framing zero \cite{Bianchi:2014laa} reads
\begin{equation}\label{label9}
\braket{\mathcal{W}_B(\nu)}_0=1+\frac{\pi ^2}{6} \left(3 \nu ^2+2\right) \lambda^2+O(\lambda^4)\,.
\end{equation}
Considering the first three order expansion of the bosonic maximal circle Wilson loop for arbitrary framing and winding given in \cite{Bianchi:2016gpg}, we formulate the following ansatz
\begin{equation}\label{label10}
\Phi_B(\nu,\nu)= \pi\nu \lambda+ \pi^3 \nu (a \nu^2+b)\lambda^3+O(\lambda^4)
\end{equation}
where $a$ and $b$ are two unknown parameters. We can uplift \eqref{label9} to generic framing by means of \eqref{framingphase} as
\begin{equation}\label{label11}
\braket{\mathcal{W}_B(\nu)}_\nu
=1+i \pi    \nu \lambda+\frac{\pi^2 }{3}\lambda^2+\frac{i \pi^3}{3}   \nu   \left((3 a+1) \nu ^2+3 b+1\right)\lambda^3+O(\lambda ^4)\,.
\end{equation}
The second equality in \eqref{label8} must reproduce the genuine calculation of the first orders of $B^\varphi_{1/6}$
\begin{equation}
\label{label12}
\frac{1}{4}\lambda^2+\frac{1}{12} \pi^2 (12a+6 b+1)\lambda^4
\overset{!}{=}
\frac{1}{4}\lambda^2-\frac{\pi^2}{4}\lambda^4
\end{equation}
as well as the expansion of circular matrix model for $\nu=1$ \cite{Marino:2009jd}\footnote{The expansion is readily available in appendix E of \cite{Bianchi:2014laa} after setting $m=1$.}
\begin{equation}
\label{label13}
1+i \pi \lambda+\frac{\pi^2 }{3}\lambda^2+\frac{i \pi^3}{3} (3 a+3 b+2)\lambda^3
\overset{!}{=}
1+i \pi \lambda +\frac{\pi^2}{3}\lambda^2+\frac{i\pi^3}{6}\lambda^3\,.
\end{equation}
These fix the free parameters $a=-1/6$ and $b=-1/3$. We conclude with a summary of what is found:
\begin{equation}\label{label14}
\begin{split}
\Phi_B(\nu,\nu)&= \pi\nu \lambda- \frac{\pi^3}{6} \nu (\nu^2+2)\lambda^3+O(\lambda^4)\\
\braket{\mathcal{W}_{B}(\nu)}_{\nu}&=
1+i \pi  \lambda  \nu +\frac{\pi ^2 \lambda^2}{3}+\frac{1}{6} i \pi ^3 \lambda ^3 \nu^3+O(\lambda ^4)\,.
\end{split}
\end{equation}

\section{Exact expression for $B_{1/2}$ and comments on $h(\lambda)$}
\label{matrix_model}
In this section we derive the non-perturbative expression of the $\frac12$BPS Bremsstrahlung function \eqref{B_half_4} from the localization result of supersymmetric Wilson loops in ABJM with purely bosonic couplings. To do this, we find convenient to organize some useful formulas that seem scattered in different papers. We use them to rewrite the $n$-wound $\frac16$BPS Wilson loop in a recursive form that takes inspiration from \cite{Klemm:2012ii}, but with the advantage that the expectation value is readily in a quite explicit function of the coupling and no differential relations are involved at all. Although not technically necessary, we quote only large-$N$ for simplicity. The result \eqref{single_winding} is naturally written in terms of the ``localization"  coupling constant $\kappa=\kappa(\lambda)$ that has a close relation to the conjectured form of the interpolating function $h(\lambda)$, which plays the role of the effective coupling absorbing the dependence on $\lambda$ in all integrability-based calculations in ABJM. This connection enable us to make few comments on the (still lacking) proof of the proposal of \cite{Gromov:2014eha} for the exact expression of $h(\lambda)$.

\subsection{$B_{1/2}$ from $\frac16$BPS Wilson loops}

Localization techniques on $S^3$ computes the expectation value of the $\frac16$BPS Wilson loop that winds $n$-times around a circle in terms of a matrix model \cite{Kapustin:2009kz,Marino:2009jd,Drukker:2010nc}. At large $N$ it is solvable in the integral form \cite{Marino:2009jd}\footnote{Notice a different Wilson loop normalization when comparing with \cite{Klemm:2012ii}.}
\begin{equation}
\label{mm1}
\langle\mathcal{W}_B^{n}(1)\rangle_1=-\frac{i}{2\pi^2\lambda} \int_{1/a}^{{a} }dx \,  e^{nx}\arctan{\sqrt{\frac{\alpha-2\cosh x}{\beta+2\cosh x}}}\,.
\end{equation}
Few comments are in order. We see shortly that the integral is not naturally a function of $\lambda$ but instead of $\kappa>0$, as defined from the latter via inversion of an hypergeometric function
\begin{equation}
\label{kappa}
\lambda=\frac{\kappa}{8\pi}\,\,{}_3F_2\left(\frac12,\frac12,\frac12;1,\frac32;-\frac{\kappa^2}{16}\right)\,.
\end{equation}
The weakly-coupled region $\lambda\ll1$ (strongly-coupled region $\lambda\gg1$) corresponds to $\kappa\ll1$ ($\kappa\gg1$). The relation can be solved perturbatively in both regimes and it represents one example of ``mirror maps'' in \cite{Marino:2009jd} between a ``bare'' coordinate ($\kappa$), in which the Wilson loop is naturally given, and a ``flat'' coordinate ($\lambda$) computed in terms of the former by a certain period integral. The integrand \eqref{mm1} is defined by the auxiliary variables
\begin{equation}
\label{alphabeta}
\alpha= 2+i \kappa\,,
\qquad
\beta= 2-i \kappa\,.
\end{equation}
The parameter $a$ and $b$, which determine the positions of the cuts (in the relevant lens space matrix model \cite{Marino:2002fk}) where the eigenvalues tend to condense in the planar limit, are
\begin{equation}
\label{ab}
a=\frac{1}{2}\left(\alpha+\sqrt{\alpha^2-4}\right)\,,
\qquad
b=\frac{1}{2}\left(\beta+\sqrt{\beta^2-4}\right)\,,
\end{equation}
with the properties
\begin{gather}
 \alpha=a+\frac{1}{a}\,,~~~~~~~
 \beta=b+\frac{1}{b}\,,~~~~~~~
 \beta=\alpha^*\,,~~~~~~~
 b=a^*\,.
\end{gather}
We quote only the integral \eqref{mm1} for the  Wilson loop associated to $U(N)_{k}$. The Wilson loop for $U(N)_{-k}$ is obtained from this by swapping $a\leftrightarrow b$ and changing the overall sign
\begin{equation}\label{mm0}
\langle\hat{\mathcal{W}}^{n}_B(1)\rangle_1
=\langle\mathcal{W}^{n}_B(1)\rangle_1^*
=-\langle\mathcal{W}^{n}_B(1)\rangle_1 \Big|_{a\leftrightarrow b}\,.
\end{equation}

Let us bring the integral \eqref{mm1} into a more explicit form. The change of variable $y=e^x$ transforms \eqref{mm1} into
\begin{equation}\label{arctan}
\langle\mathcal{W}_B^{n}(1)\rangle_1
=-\frac{i}{2\pi^2\lambda}  \int_{1/a}^a dy\,  y^{n-1}\arctan \sqrt{\frac{-y^2+\alpha y -1}{y^2+\beta y +1}}
\end{equation}
and partial integration eliminates the trigonometric function in the integrand\footnote{We neglected $y$-dependent complex phases to put the denominator under a unique square root. However, this is eventually harmless because the difference of $\mathcal{I}$'s in the last equality of \eqref{mm2}, computed as in \eqref{I_result}, agrees with \eqref{arctan} numerically.}
\begin{equation}\label{mm2}
\langle\mathcal{W}_B^{n}(1)\rangle_1
= \frac{i}{4\pi^2 n \lambda}  \int_{1/a}^a dy \frac{y^{n+1}-y^{n-1}}{\sqrt{(-y^2+\alpha y -1)(y^2+\beta y +1)}}
=\frac{i}{2\pi^2 n \lambda}  (\mathcal{I}_{n+1}-\mathcal{I}_{n-1})\,,
\end{equation}
where one defines \cite{Klemm:2012ii}
\begin{equation}\label{I_definition}
\mathcal{I}_n \equiv \int_{1/a}^a \frac{y^n}{2\sqrt{(-y^2+\alpha y -1)(y^2+\beta y +1)}}\,,\qquad n\in\mathbb{N}\,.
\end{equation}
The solution comes from expanding the ``generating functional'' \cite{gradshteyn}
\begin{flalign}
\label{gen_fun}
& \int_{1/a}^a \frac{dy}{(p-y)\sqrt{(-y^2+\alpha y -1)(y^2+\beta y +1)}}
 = \frac{2\sqrt{a}\sqrt{b}}{(a+b)(b+p)(ap-1)}\\
&~~~~~~~\times \left[(1-ap)\mathbb{K}\left(-\frac{(a^2-1)(b^2-1)}{(a+b)^2}\right)-(1+ab)\Pi\left(\frac{(a^2-1)(b+p)}{(a+b)(ap-1)}\right)\right]\nonumber
\end{flalign}
for $p\to\infty$ and equating powers of $p^{-1}$. The result is \cite{Klemm:2012ii}\footnote{An error in this paper is corrected.}
\begin{equation}\label{I_result}
\mathcal{I}_n = \frac{\sqrt{ab}}{1+ab} \sum_{j=0}^n {{n}\choose{j}} (-1)^{n+j+1} (a+b)^j b^{n-j} V_j
\end{equation}
where $V_j$ is given recursively in terms of Jacobi elliptic functions\footnote{
In the Abramowitz \& Stegun/Mathematica notation, e.g. appendix A in \cite{Vescovi:2016zzu}: $E$ and $\Pi$ are the incomplete elliptic integrals of the second and third kind, $\mathbb{K}$ and $\mathbb{E}$ are the complete elliptic integrals of the first and second kind.}
\begin{equation}
\label{Vj}
\begin{split}
V_0&=\mathbb{K}(s)\,, \qquad
V_1=\Pi(t|s)\,,\\
V_2&=\frac{1}{2 (t-1)(s-t)} \left[
t \mathbb{E}(s)+(s -t)\mathbb{K}(s)+(2 st+2 t-t^2-3s)\Pi(t|s)
\right]\,,\\
V_{m+3}&=\frac{1}{2(m+2)(t-1)(s-t)}\left[
-(2m+1)s V_m - 2(m+1) (st+t-3s) V_{m+1}\right. \\
& \left.
-(2m+3)(t^2-2st-2t+3s)V_{m+2} 
\right]\,,
\qquad\qquad m\geq0\,, 
\end{split}
\end{equation}
with
\begin{equation}
\label{Vj2}
s=\frac{(a^2-1)(b^2-1)}{(1+ab)^2}\,,
\qquad\qquad
t=\frac{1-a^2}{1+ab}\,.
\end{equation}
Formula \eqref{mm2} together with \eqref{alphabeta}-\eqref{ab} and \eqref{I_result}-\eqref{Vj2} expresses the expectation value \eqref{mm1} as a function of $\kappa$, and so of $\lambda$ via the inversion of \eqref{kappa}.

\begin{figure}[t]
  \begin{center}
  \includegraphics[width=6.5cm]{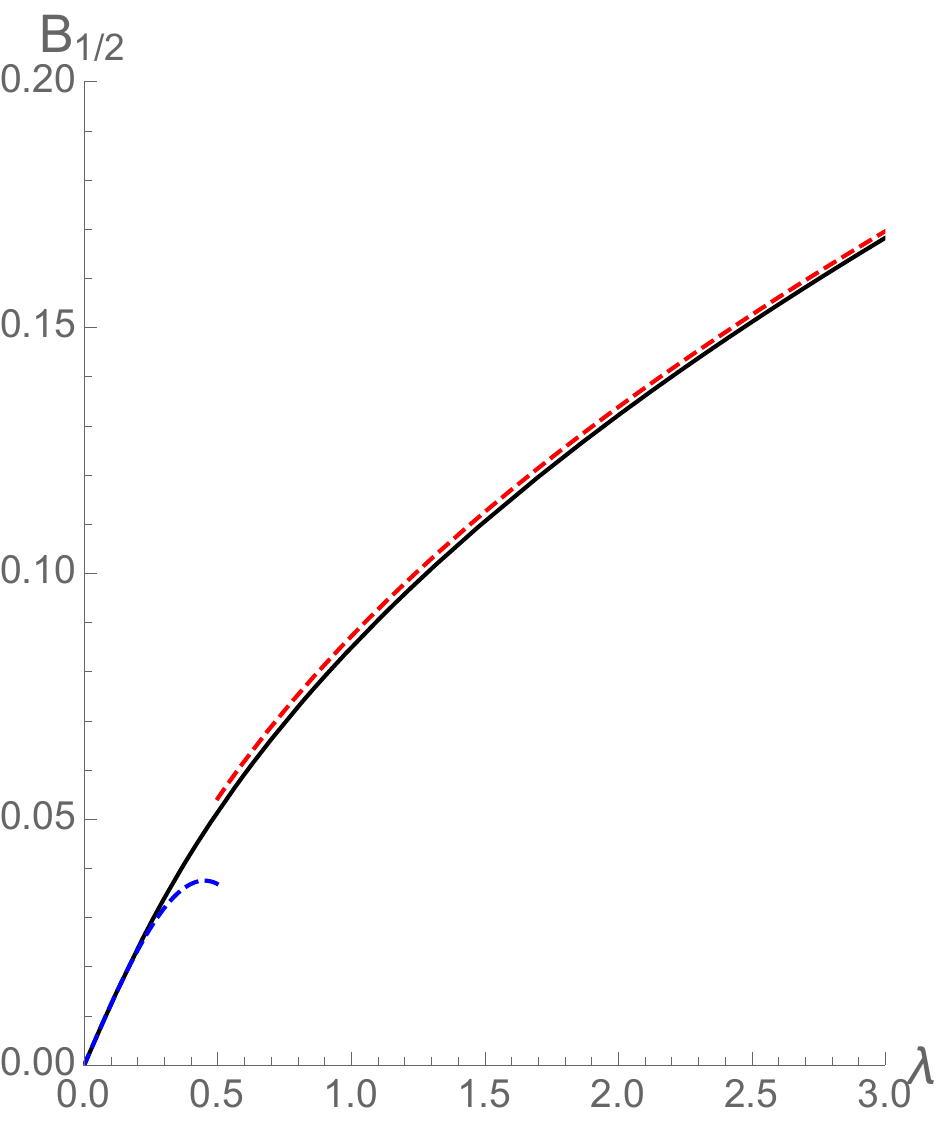}
\hfill
  \includegraphics[width=6.5cm]{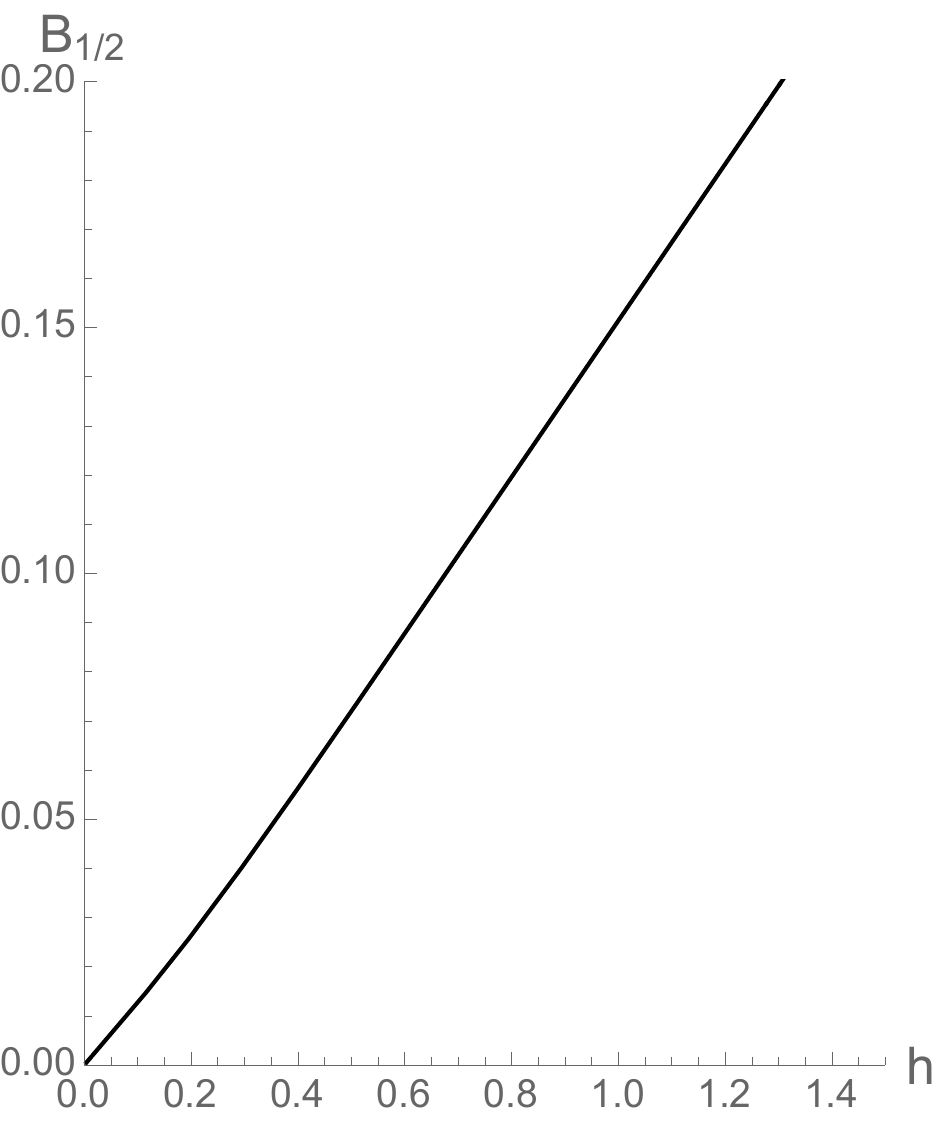}
  \caption{Left panel: plot of \eqref{B_exact} (black curve) and its weak/strong coupling expansions \eqref{B_exact_pert1} and \eqref{B_exact_pert2} (blue/red  curves) as functions of the 't Hooft coupling. Right panel: plot of \eqref{B_exact} as function of the interpolating function of ABJM integrability.}
  \label{fig:B}
  \end{center}
\end{figure}

We find it hard to write the $n$th-derivative of \eqref{mm2}\footnote{An alternative result is available at any winding number and finite $N$ (neglecting exponentially small corrections) in terms of Airy functions \cite{Klemm:2012ii}. However it does not seem to us that it is possible to expand it consistently at $\lambda\ll 1$ after the planar limit is imposed, which justifies why we choose not to make use of it for the purpose of the derivative.} at $n=1$, thus preventing a compact formula for the $\frac16$BPS Bremsstrahlung functions \eqref{brem16}. To this aim, what one ideally needs is a solution of \eqref{I_definition} for real $n$, which we could not obtain relying on \eqref{gen_fun}. Alternatively, one can try to solve the recurrence \eqref{Vj} in closed form and then promote the index of $V_j$ to be continuous, but this is not very useful either because of the discrete sum in \eqref{I_result}. The case of single-wound Wilson loop is instead very simple
\begin{flalign}
\label{single_winding}
&\langle\mathcal{W}_B^{1}(1)\rangle_1
=\frac{i}{4 \pi ^2\lambda \sqrt{ab} (1+a b)}\left[b (a^2 b-ab^2+3a+b) \,\mathbb{K}\!\left(\frac{\left(a^2-1\right) \left(b^2-1\right)}{(a b+1)^2}\right)\right.\\
&~~\left.-(1+a b)^2 \,\mathbb{E}\!\left(\frac{\left(a^2-1\right) \left(b^2-1\right)}{(1+a b)^2}\right)+(a^2-b^2) (1-a b) \,\Pi\! \left(\frac{1-a^2}{1+a b}\left|\frac{\left(a^2-1\right) \left(b^2-1\right)}{(1+a b)^2}\right.\right)\right]\,.\nonumber
\end{flalign}
The imaginary part \eqref{tanPhi} is easily extracted with a property that gives the elliptic integral $\Pi$ when $a$ and $b$ are swapped (see above \eqref{mm0})
\begin{flalign}
& \Pi\! \left(\frac{1-b^2}{1+a b}\left|\frac{\left(a^2-1\right) \left(b^2-1\right)}{(1+a b)^2}\right.\right) = 
-\Pi\! \left(\frac{1-a^2}{1+a b}\left|\frac{\left(a^2-1\right) \left(b^2-1\right)}{(1+a b)^2}\right.\right) \\
&\qquad\qquad\qquad\qquad\qquad
+\frac{\pi(1+ab)}{2\sqrt{ab}(a+b)}
+\mathbb{K}\!\left(\frac{\left(a^2-1\right) \left(b^2-1\right)}{(a b+1)^2}\right)\,.\nonumber
\end{flalign}
This leads to the desired expression for the Bremsstrahlung function \eqref{B_half_4}
\begin{flalign}\label{B_exact}
B_{1/2}&=
\frac{i}{8\pi}\left[1+\frac{4 \sqrt{a} b^{3/2} (3a+b+a^2b-ab^2) }{\pi (a-b) (a^2 b^2-1)} \mathbb{K}\left(\frac{\left(a^2-1\right) \left(b^2-1\right)}{(a b+1)^2}\right)    -\frac{ 4 \sqrt{ab} (1+a b)}{ \pi (a-b) (a b-1)}\right.  \nonumber\\
&\left. \times \mathbb{E}\left(\frac{\left(a^2-1\right) \left(b^2-1\right)}{(1+a b)^2}\right)
-\frac{4 \sqrt{ab} (a+b) }{\pi (1+a b)}\Pi \left(\frac{1-a^2}{1+a b}\left|\frac{\left(a^2-1\right) \left(b^2-1\right)}{(1+a b)^2}\right.\right) \right]\,,
\end{flalign}
where we remind that
\begin{gather}
a=1+\frac{i}{2}\kappa+\frac{i}{2}\sqrt{\kappa(\kappa-4i)}\,,
\qquad\qquad
b=1-\frac{i}{2}\kappa-\frac{i}{2}\sqrt{\kappa(\kappa+4i)}
\end{gather}
are functions of $\lambda$ through \eqref{kappa}. The parameters are complex but \eqref{B_exact} is real by construction \eqref{B_half_4}, see also left panel in Figure \ref{fig:B}, although not manifestly. 

As a check, we perform a weak coupling expansion. In the region $\lambda\ll1$ the inversion of \eqref{kappa} delivers \cite{Marino:2009jd}
\begin{gather}
\kappa=8\pi\lambda+\frac{8\pi^3}{3}\lambda^3-\frac{14\pi^5}{15}\lambda^5+\frac{346\pi^7}{315}\lambda^7+O(\lambda^9)
\end{gather}
and one obtains
\begin{gather}\label{B_exact_pert1}
B_{1/2}=\frac{\lambda}{8}-\frac{\pi^2 \lambda^3}{48}+O(\lambda^5) \,.
\end{gather}
It reproduces the known coefficients in calculations up to three loops \cite{Griguolo:2012iq,Bianchi:2014laa,Bianchi:2017svd}. In the opposite regime $\lambda\gg1$ one has \cite{Marino:2009jd}
\begin{gather}
\kappa=e^{\pi\sqrt{2\lambda}}\big(1+O(\lambda^{-1/2},e^{-2\pi\sqrt{2\lambda}})\big)
\end{gather}
and from\footnote{We expand the elliptic integral for the two arguments going to 1 and then expand in $\kappa$.}
\begin{gather}
\Pi \left(\frac{1-a^2}{1+a b}\left|\frac{\left(a^2-1\right) \left(b^2-1\right)}{(1+a b)^2}\right.\right) = -\frac{3i}{8}\kappa\log\kappa+\frac{\pi}{16}\kappa+O(\log\kappa)\,,
\qquad
\kappa\to\infty
\end{gather}
it follows that
\begin{gather}\label{B_exact_pert2}
B_{1/2}=\frac{\sqrt{\lambda}}{2\sqrt{2}\pi}-\frac{1}{4\pi^2}+O(\lambda^{-1/2})\,, \qquad \lambda\gg 1\,.
\end{gather}
This agrees with the classical and one-loop order in calculations in string theory \cite{Forini:2012bb,Correa:2014aga,Aguilera-Damia:2014bqa}.

\subsection{Interpolating function}
A conjecture for the exact expression of the interpolating function of ABJM integrability was put forward in \cite{Gromov:2014eha}. The proposal takes the form of an implicit equation
\begin{equation}
\label{h}
\lambda=\frac{\sinh(2\pi h(\lambda))}{2\pi}{}_3F_2\left(\frac12,\frac12,\frac12;1,\frac32;-\sinh^2(2\pi h(\lambda))\right)
\end{equation}
based on the similarity between two exact results in ABJM: the slope function \cite{Basso:2011rs} describing the small-spin limit of $SL(2)$ operators as derived via integrability \cite{Cavaglia:2014exa} and the $\frac16$BPS Wilson loop \eqref{mm1} via localization. In particular, one recognizes that $h(\lambda)$ should have a very simple expression in terms of the localization effective coupling \eqref{kappa}
\begin{equation}\label{kappa_h}
\kappa(\lambda) = 4\sinh(2\pi h(\lambda))
\end{equation}
where the 't Hooft coupling is absorbed in these two ``couplings" and does not appear explicitly. As suggested in \cite{Gromov:2014eha}, a rigorous derivation of \eqref{h} would require the comparison of the Bremsstrahlung function obtained as function of $h(\lambda)$ from thermodynamic Bethe ansatz equations \cite{Bombardelli:2009xz,Gromov:2009at} or quantum spectral curve method \cite{Bombardelli:2017vhk} and the localization prediction \eqref{B_exact} function of $\kappa(\lambda)$, paralleling a similar derivation done in $\mathcal{N}=4$ SYM \cite{Correa:2012hh}. The comparison has not been done yet in ABJM because the integrability calculation is still lacking at the moment. Here we want instead to assume that \eqref{h} is correct and use the comparison to make an explicit prediction on the result of the integrability calculation of $B_{1/2}$. One should take \eqref{B_exact} with $a,b$ given by \eqref{alphabeta}-\eqref{ab} in terms of $\kappa$ and then trade $\kappa$ with the interpolating function using \eqref{kappa_h}. This yields $B_{1/2}$ as an explicit function of $h(\lambda)$ (Figure \ref{fig:B}, right panel) that will be very interesting to derive from first principles in the future.

\subsection*{Acknowledgements}
We thank Marco Bianchi, Jo\~ao Caetano, Martina Cornagliotto, Luca Griguolo, Vladimir Kazakov, Madalena Lemos, Fedor Levkovich-Maslyuk, Andrea Mauri, Marco Meineri, Silvia Penati, Domenico Seminara and Diego Trancanelli for discussions and the authors of \cite{Bianchi:2018bke} for sharing with us their results prior publication. We are particularly grateful to Luca Griguolo and Domenico Seminara for valuable help in the initial stage of the project. The work of LB is supported by Deutsche Forschungsgemeinschaft in Sonderforschungsbereich 676 ``Particles, Strings, and the Early Universe" and by the European Union’s Horizon 2020 research and innovation
programme under the Marie Sklodowska-Curie grant agreement No 749909. The work of MP is supported by ``Della Riccia Foundation" grant and by the European Research Council (Programme ``Ideas'' ERC-2012-AdG 320769 AdS-CFT-solvable). EV acknowledges financial support by FAPESP grants 2014/18634-9 and 2016/09266-1 and he thanks Yunfeng Jiang and ETH Zurich for hospitality while this project was in preparation.

\appendix
\section{Conventions}
The $\frac16$BPS Wilson line breaks the $SU(4)$ R-symmetry down to $SU(2)\times SU(2)\times U(1)$. In particular the $SU(4)$ fundamental indices are split as follows
\begin{equation}
 I=(a, \dot a)
\end{equation}
where $a$ is a fundamental index of the first $SU(2)$ factor and $\dot a$ of the second one. Those are raised and lowered by the action of epsilon tensors with the following conventions:
\begin{align}
\e^{12}=\e^{\dot{1} \dot{2}}&=1 & \e_{12}&=\e_{\dot{1} \dot{2}}=-1
\end{align}
such that
\begin{align}
 \e^{ab}\e_{bc}&=\d^a_c  &  \e^{\dot a \dot b}\e_{\dot b \dot c}&=\d^{\dot a}_{\dot c}\,.
\end{align}
The spinor contractions and the conventions of the raising and lowering spinorial indices are as follows
\begin{equation}
 \chi \bar \chi\equiv \chi^{\a} \bar \chi_{\a} \qquad \chi^{\a}=\e^{\a\b}\chi_{\b} \qquad \e^{\a\b} \e_{\b \g}=\d^\a_{\g} \qquad \e^{12}=-\e_{12}=1\,.
\end{equation}
We work in the Euclidean space parametrized by the vector $x^\mu=\{x^1,x^2,x^3\}$ with $\gamma$ matrices 
\begin{equation}
 {(\gamma^{\mu})_{\a}}^{\b}=(\s^1,\s^2,-\s^3)
\end{equation}
satisfying the Clifford algebra $\{\g^{\mu},\g^{\nu}\}=2\d^{\mu\nu} \mathbb{1}$ and obeying the following relation
\begin{equation}
{(\gamma^{\mu})^{\a}}_{\b}=\e^{\a\g}\e_{\b\d}{(\gamma^{\mu})_{\g}}^{\d}\quad\Longrightarrow\quad {(\gamma^{\mu})^{\a}}_{\b}=(\s^1,-\s^2,-\s^3)
\end{equation}
where $\sigma$'s are the Pauli matrices. When gamma matrices are involved and no spinor indices are specified the following convention is assumed
\begin{equation}
 \chi \gamma^{\mu} \bar\chi=\chi^{\a} {(\gamma^{\mu})_{\a}}^{\b} \bar \chi_{\b}\,.
\end{equation}
In this paper for the straight line we use the $\pm$ basis, then it is useful to write down the gamma matrices projected in this basis
\begin{equation}
 {(\gamma^{\mu}_{\pm \text{ basis}})_\pm}^{\pm}=\{\s_3,\s_2,-\s_1\}\,.
\end{equation}

\section{Bremsstrahlung functions and operator insertions}
\label{app:correa}
In this section we justify and motivate the introduction of an absolute value in the prescription for computing $B_{1/6}^{\theta}$ in terms of the latitude Wilson loops. All expectation values in this appendix are at framing 0, hence the subscript is suppressed. Consider first the generalized cusp described in section \ref{gencusp}. When taking a double derivative with respect to the internal angle $\theta$, in our notation we get a the following combination of defect two-point functions
\begin{equation}
 \left.\frac{\partial^2}{\partial \theta^2} \log \braket{\mathcal{W}_B[C]}\right|_{\theta=0}=\frac14 \int_{0}^{\infty} d\tau_1\int_{\tau_1}^{\infty} d\tau_2\left(\braket{\mathbb{O}^{2\dot 1}(\tau_1) \bar{\mathbb{O}}^{\dot 2 1}(\tau_2)}_{\mathcal{W}} +\braket{ \bar{\mathbb{O}}^{\dot 2 1}(\tau_1)\mathbb{O}^{2\dot 1}(\tau_2)}_{\mathcal{W}}\right)
\end{equation}
where one should keep in mind that $\mathbb{O}^{2 \dot 1}=\frac{4\pi i}{k} C_1 \bar C^3$. Although the symmetry fixes
\begin{equation}
  \braket{\mathbb{O}^{2\dot 1}(\tau_1) \bar{\mathbb{O}}^{\dot 2 1}(\tau_2)}_{\mathcal{W}} =\braket{ \bar{\mathbb{O}}^{\dot 2 1}(\tau_1)\mathbb{O}^{2\dot 1}(\tau_2)}_{\mathcal{W}}
\end{equation}
let us suppose for a second that they are determined by the general form \eqref{2ptfunOO} with two different coefficients $c_s$ and $\bar c_s$. Therefore we get
\begin{equation}
    \left.\frac{\partial^2}{\partial \theta^2} \log \braket{\mathcal{W}_B[C]}\right|_{\theta=0}=-\frac{c_s+\bar{c}_s}4 \int_{0}^{\infty} d\tau_1\int_{\tau_1}^{\infty} d\tau_2\frac{1}{|\tau_{12}|^2}\,.
\end{equation}
Switching to the cylinder parametrization (see \cite{Aguilera-Damia:2014bqa,Bianchi:2017ozk}) we get
\begin{equation}
   \left.\frac{\partial^2}{\partial \theta^2} \log \braket{\mathcal{W}_B[C]}\right|_{\theta=0}=-\frac{c_s+\bar{c}_s}4 \int_{-\infty}^{\infty} d\tau_1\int_{\tau_1}^{\infty} d\tau_2\frac{1}{2(\cosh \tau_{12}-1)}=T \ \frac{c_s+\bar{c}_s}{4}
\end{equation}
 where $T$ is an overall divergence corresponding to the integral $\int_{-\infty}^{\infty}d\tau$. From that, using the same arguments of \cite{Aguilera-Damia:2014bqa,Bianchi:2017ozk} we can write
 \begin{equation}\label{Bline}
  B^\theta_{1/6}=-\frac1{2T}  \left.\frac{\partial^2}{\partial \theta^2} \log \braket{\mathcal{W}_B[C]}\right|_{\theta=0}=-\frac{c_s+\bar{c}_s}{8}\,.
 \end{equation}

On the other hand, when we consider the derivative of the latitude Wilson loop of section \ref{sec:2.1} we have
\begin{multline}
 \left.\frac{\partial^2 \log \braket{\mathcal{W}_B(\nu)}}{\partial \theta_0^2}\right|_{\theta_0=0}\\=-\frac14 \int_{0}^{2\pi} d\tau_1\int_{\tau_1}^{2\pi} d\tau_2\left(e^{i\t_{12}}\braket{\mathbb{O}^{2\dot 1}(\tau_1) \bar{\mathbb{O}}^{\dot 2 1}(\tau_2)}_{\mathcal{W}} +e^{-i\t_{12}}\braket{ \bar{\mathbb{O}}^{\dot 2 1}(\tau_1)\mathbb{O}^{2\dot 1}(\tau_2)}_{\mathcal{W}}\right)
\end{multline}
which gives
\begin{equation}
 \left.\frac{\partial^2 \log \braket{\mathcal{W}_B(\nu)}}{\partial \theta_0^2}\right|_{\theta_0=0}=-\frac{\pi}{2} \int_{0}^{2\pi} d\tau_1\int_{\tau_1}^{2\pi} d\tau_2\left(\frac{e^{i\t_{12} }c_s + e^{-i \tau_{12} }\bar{c}_s}{2(1-\cos \tau_{12})}\right)\,.
\end{equation}
This gives a real and an imaginary contribution
\begin{equation}\label{anom}
 \left.\frac{\partial^2 \log \braket{\mathcal{W}_B(\nu)}}{\partial \theta_0^2}\right|_{\theta_0=0}=-\frac{\pi}{2} \int_{0}^{2\pi} d\tau_1\int_{\tau_1}^{2\pi} d\tau_2\left(\frac{(c_s+\bar{c}_s) \cos \t_{12} + i (c_s - \bar{c}_s) \sin \tau_{12}}{2(1-\cos \tau_{12})}\right)\,.
\end{equation}
As we mentioned $c_s=\bar c_s$, but since $(c_s-\bar c_s)$ multiplies a divergent integral the overlapping of these effects may produce an anomaly-like contribution in equation \eqref{anom}. The perturbative computation of \cite{Bianchi:2018bke} actually shows that this is indeed the case. To avoid this issue and recover the expression for the Bremsstrahlung function in equation \eqref{Bline} we combine \eqref{anom} with its complex conjugate, so that the imaginary part disappears
\begin{equation}
 \frac12\left(\left.\frac{\partial^2 \log \braket{\mathcal{W}_B(\nu)}}{\partial \theta_0^2}\right|_{\theta_0=0}+ \left.\frac{\partial^2 \log \braket{\hat{\mathcal{W}}_B(\nu)}}{\partial \theta_0^2}\right|_{\theta_0=0}\right)=-\frac{\pi^2}{2} (c_s+\bar c_s)=-\pi^2 c_s\,.
\end{equation}
Therefore, reinstating the subscript for framing 0 we conclude
\begin{equation}\label{B_sixth_theta_3}
B^{\theta}_{1/6}=\frac{1}{4\pi^2} \Big. \partial_\nu \log |\langle \mathcal{W}_B (\nu) \rangle_0| \Big|_{\nu=1}\,.
\end{equation}

\section{The $su(1,1|1)$ subalgebra}\label{subalgebra}
We start from the $osp(6|4)$ supersymmetry algebra, which contains the three-dimensional conformal algebra
\begin{align}
[P^{\mu},K^{\nu}]&=-2\d^{\mu\nu} D-2 M^{\mu\nu} & [D,P^{\mu}]&= P^{\mu} & [D,K^{\mu}]&=-K^{\mu}\\
 [M^{\mu\nu}, M^{\rho\sigma}]&=\d^{\sigma[\mu} M^{\nu]\rho}+\d^{\rho[\nu} M^{\mu]\sigma} & [P^{\mu},M^{\nu\rho}]&=\delta^{\mu[\nu}P^{\rho]} & [K^{\mu},M^{\nu\rho}]&=\delta^{\mu[\nu}K^{\rho]} \,,
\end{align}
the $SU(4)$ generators
\begin{align}
 [{J_I}^J,{J_K}^L]=\d_I^L {J_K}^J-\d^J_{K} {J_I}^L
\end{align}
and the fermionic generators $Q^{IJ}_\a$ and $S^{IJ}_\a$  
\begin{align}
 \{Q^{IJ}_\a,Q^{KL\b}\}&=2\e^{IJKL} {(\gamma^\mu)_{\a}}^{\b} P_\mu \qquad  \{S^{IJ}_\a,S^{KL\b}\}=2\e^{IJKL} {(\gamma^\mu)_{\a}}^{\b} K_\mu \label{genanticomm}\\  \{Q^{IJ}_\a,S^{KL\b}\}&=i\e^{IJKL} ({(\gamma^{\mu\nu})_{\a}}^{\b} M_{\mu\nu}+2\d_{\a}^{\b} D)+2i\d_{\a}^{\b}\e^{KLMN}(\d_M^J {J_N}^I-\d_M^I {J_{N}}^J)
\end{align}
with the reality condition $\bar Q_{IJ\a}=\frac12 \e_{IJKL} Q^{KL}_{\a}$. Mixed commutators are
\begin{align}
 [D,Q^{IJ}_{\a}]&=\frac12 Q^{IJ}_{\a} & [D,S^{IJ}_{\a}]&=-\frac12 S^{IJ}_{\a} \\
 [M^{\mu\nu},Q^{IJ}_{\a}]&=-\frac12 {(\gamma^{\mu\nu})_{\a}}^{\b} Q^{IJ}_{\b} & [M^{\mu\nu},S^{IJ}_{\a}]&=-\frac12 {(\gamma^{\mu\nu})_{\a}}^{\b} S^{IJ}_{\b}  \\
 [K^{\mu},Q_{\a}^{IJ}]&=i{(\g^{\mu})_{\a}}^{\b} S^{IJ}_{\b} & [P^{\mu},S^{IJ}_{\a}]&=i{(\g^{\mu})_{\a}}^{\b} Q^{IJ}_{\b}\\
 [{J_I}^J,Q^{KL}_{\a}]&=\d_I^K Q^{JL}_\a+\d_{I}^L Q^{KJ}_\a-\frac12\d_I^J Q^{KL}_{\a} &  [{J_I}^J,S^{KL}_{\a}]&=\d_I^K S^{JL}_\a+\d_{I}^L S^{KJ}_\a-\frac12\d_I^J S^{KL}_{\a}\,.
\end{align}

The $su(1,1|1)$ subalgebra generated by $\{D,P,K,J;Q,\bar Q, S, \bar S\}$ is preserved by the $\frac16$BPS Wilson line, with the identifications
\begin{align}
 P&\equiv P_1 & K&\equiv K_1 & J&\equiv {J_1}^1+{J_2}^2-i M^{23}\\
 Q&\equiv Q^{12}_+ & \bar Q&\equiv Q^{34}_-  & S&\equiv -iS^{12}_+ & \bar S&\equiv iS^{34}_-\,.
\end{align}
The $su(1,1)$ commutation relations are
\begin{align}\label{comm1}
 [P,K]&=-2 D & [D,P]&=P & [D,K]&=-K\,.
\end{align}
The anticommutation relations for the fermionic generators read
\begin{align}
 \{Q,\bar Q\}&= 2P &   \{S,\bar S\}&= 2K \label{comm2}\\
 \{Q, \bar S\}&=2 D- 2  J & \{S,\bar Q\}&= 2 D+2J  \label{comm3}
\end{align}
and mixed commutators
\begin{align}
 [D,Q]&=\frac12 Q  & [D,\bar Q]&=\frac12 \bar Q & [D,S]&=-\frac12 S &  [D,\bar S]&=-\frac12 \bar S\label{comm4}\\
 [K,Q]&= S & [K,\bar Q]&= \bar S & [P,S]&=-Q & [P,\bar S]&=- \bar Q\label{comm5}\\  
 [J,Q]&=\frac12 Q & [J,\bar Q]&=-\frac12 \bar Q & [J,S]&=\frac12 S & [J,\bar S]&=-\frac12 \bar S\,. \label{comm6}
\end{align}

\subsection{Representations of $su(1,1|1)$}
We briefly review the classification of long and short multiplets of $su(1,1|1)$. The algebra is characterized by two Dynkin labels $[\D,j]$ associated to the Cartan generators of the subalgebra $su(1,1)\oplus u(1)$. The supercharges carry the following charges
\begin{align}
 Q &\quad [\tfrac12,\tfrac12] & \bar Q &\quad [\tfrac12,-\tfrac12] & S &\quad [-\tfrac12,\tfrac12] & \bar S &\quad [-\tfrac12,\tfrac12]\,.
\end{align}
A highest weight state $\ket{\D,j}$ is defined by the condition
\begin{equation}
 S\ket{\D,j}=\bar S\ket{\D,j}=0
\end{equation}
and a long multiplet $\mathcal{A}_{\D,j}$ can be easily built by acting on it with the supercharges $Q$ and $\bar Q$
\begin{center}
\begin{tikzpicture}
  \node[above] at (0,4.9) {$[\D,j]$};
  \draw[->]  (0.35,4.9)--(0.65,4.6);
  \node[above] at (1.5,3.9) {$[\D+\tfrac12,j+\tfrac12]$};
  \node[above] at (-1.5,3.9) {$[\D+\tfrac12,j-\tfrac12]$};
  \node at (0.7,4.9) {$Q$};
  \node at (-0.7,4.9) {$\bar Q$};
  \draw[->]  (-0.35,4.9)--(-0.65,4.6);
  \draw[<-]  (0.35,3.6)--(0.65,3.9);
  \draw[<-]  (-0.35,3.6)--(-0.65,3.9);
  \node[above] at (0,2.9) {$[\D+1,j]$};
  \end{tikzpicture}
\end{center}
There are then two possible shortening conditions, generating two $\frac12$BPS multiplets
\begin{align}
j&>0 & Q\ket{\D,j}=0 \quad &\Rightarrow \quad \D=j & &\mathcal{B}_{j}\\
j&<0 & \bar Q \ket{\D,j}=0 \quad &\Rightarrow \quad \D=-j &  &\bar{\mathcal{B}}_j
\end{align}
Those multiplets contain only two operators (besides the infinite tower of conformal descendants)
\begin{center}
\begin{tikzpicture}
\node at (-0.6, 6.3) {$\mathcal{B}_j$};
  \node[above] at (0,4.9) {$[j,j]$};
  \draw[->]  (-0.35,4.9)--(-0.65,4.6);
  \node[above] at (-1.5,3.9) {$[j+\tfrac12,j-\tfrac12]$};
  \node at (-0.7,4.9) {$\bar Q$};
  \end{tikzpicture}\hspace{2 cm}
\begin{tikzpicture}
\node at (0.6, 6.3) {$\bar{\mathcal{B}}_j$};
  \node[above] at (0,4.9) {$[-j,j]$};
  \draw[->]  (0.35,4.9)--(0.65,4.6);
  \node[above] at (1.5,3.9) {$[-j+\tfrac12,j+\tfrac12]$};
  \node at (0.7,4.9) {$Q$};
  \end{tikzpicture}
\end{center}
and the long multiplet at the unitarity bound simply decomposes as
\begin{align}
 j&>0 & \mathcal{A}_{j,j}&=\mathcal{B}_j\oplus \mathcal{B}_{j+\frac12}\\
  j&<0 & \mathcal{A}_{-j,j}&=\bar{\mathcal{B}}_{j}\oplus \bar{\mathcal{B}}_{j-\frac12}\,.
\end{align}

\section{SUSY transformations}
In our conventions the $\mathcal{N}=6$ supersymmetry transformations for the fundamental fields read
\begin{align}
\d C_I&=-\theta_{IJ} \bar \psi^{J} \nonumber\\
\d \bar C^I&=- \bar \theta ^{IJ} \psi_{J}\nonumber\\
\d \psi_I^{\a}&=-2\theta_{IJ}^\b {(\g^{\mu})_\b}^{\a} D_\mu \bar C^J+\frac{4\pi}{k} \theta_{IJ}^\a (\bar C^J C_K \bar C^K-\bar C^K C_K \bar C^J)+\frac{8\pi}{k}\theta_{KL}^{\a} \bar C^K C_I \bar C^L\nonumber\\
\d \bar \psi^I_{\a}&=-2\bar \theta^{IJ\, \b} (\g^{\mu})_{\b\a} D_\mu  C_J+\frac{4\pi}{k} \bar \theta^{IJ}_\a ( C_K \bar C^K C_J-C_J\bar C^K C_K )+\frac{8\pi}{k}\bar \theta^{KL}_{\a} C_L\bar C^I C_K \nonumber\\
\d A_{\mu}&=-\frac{2\pi i}{k} (\bar \theta^{IJ} \g_{\mu} C_I \psi_J+\theta_{IJ}\g_{\mu} \bar\psi^I \bar C^J)\nonumber\\
\d \hat A_{\mu}&=-\frac{2\pi i}{k} (\bar \theta^{IJ} \g_{\mu}  \psi_J C_I+\theta_{IJ}\g_{\mu}  \bar C^J \bar\psi^I) \,.\label{N6susy}
\end{align}
The action of the supercharge $Q^{IJ}_{\a}$ can be obtained by applying of the differential operator
\begin{align}
 Q^{IJ}_{\a}=\frac{\pa}{\pa \theta_{IJ}^{\a}}
\end{align}
and using the reality condition $\bar \theta^{IJ}=\frac12 \e^{IJKL} \theta_{KL}$ with $\epsilon^{1234}=1$.

\subsection{Preserved SUSY transformations}\label{pressusy}
From the transformations \eqref{N6susy} we extract the variation of the fundamental fields under the preserved supercharges $Q$ and $\bar Q$
\begin{align}
 Q C_a&=\e_{ab} \bar \psi_+^b & Q C_{\dot a}&=0 &
 Q \bar C^a&=0 & Q \bar C^{\dot a}&= \e^{\dot a \dot b} \psi_{\dot b}^- \\
  \bar Q  C_a&=0 & \bar Q C_{\dot a}&= \e_{\dot a \dot b} \bar \psi^{\dot b}_- &
 \bar Q \bar C^a&=- \e^{ab} \psi_b^+ & \bar Q \bar C^{\dot a}&=0 \\
 Q \psi_a^+&=2 \e_{ab} \mathcal{D}_{\tau} \bar C^b &  Q \psi_a^-&=-2i \e_{ab} D \bar C^b & Q \psi_{\dot{a}}^+&=\frac{8\pi }{k} \e_{ab} \bar{C}^a C_{\dot{a}} \bar{C}^b & Q \psi_{\dot{a}}^-&=0\\
 Q \bar \psi^a_+&=0 &  Q\bar \psi^a_-&=-\frac{8\pi }{k} \e^{\dot a \dot b} {C}_{\dot a} \bar C^{a} {C}_{\dot b} & Q \bar \psi^{\dot{a}}_+&=-2i \e^{\dot a \dot b} D C_{\dot b} & Q \bar \psi^{\dot{a}}_-&=-2\e^{\dot a \dot b} \mathcal{D}_{\tau} C_{\dot b}\\
 \bar Q \psi_a^+&=0 &  \bar Q \psi_a^-&=-\frac{8\pi}{k} \e_{\dot a \dot b } \bar{C}^{\dot a} C_{a} \bar{C}^{\dot b} & \bar Q \psi_{\dot{a}}^+&=2i\e_{\dot a \dot b} \bar D \bar C^{\dot b} & \bar Q \psi_{\dot{a}}^-&=-2 \e_{\dot a \dot b} \mathcal{D}_{\tau} \bar C^{\dot b} \!\!\!\\
 \bar Q \bar \psi^a_+&=-2 \e^{ab} \mathcal{D}_{\tau} C_b & \bar Q\bar \psi^a_-&=-2i\e^{ab}\bar D C_b &\bar Q \bar \psi^{\dot{a}}_+&=\frac{8\pi}{k} \e^{ a  b} {C}_{ a} \bar C^{\dot a} {C}_{ b} &\bar Q \bar \psi^{\dot{a}}_-&=0
 \end{align}
 \begin{align}
 Q A_1&=\frac{2\pi i}{k}(\e^{\dot a \dot b} C_{\dot a} \psi_{\dot b}^- + \e_{ab} \bar \psi_+^a \bar C^b ) &  \bar Q A_1&=\frac{2\pi i}{k}(-\e_{\dot a \dot b}  \bar \psi^{\dot a}_- \bar C^{\dot b}+ \e^{ab}   C_a \psi^+_b)
\end{align}
where 
\begin{align}
 \mathcal{D}_{\tau} C_a&= \pa_{\tau} C_a+ i \mathcal{L}_B C_a- i C_a \hat{\mathcal{L}}_B & \mathcal{D}_{\tau} \bar C^a&= \pa_{\tau} \bar C^a+ i \hat{\mathcal{L}}_B \bar C^a- i \bar C^a {\mathcal{L}_B} \\
 D&=D_2-i D_3 & \bar D&=D_2+i D_3\,.
\end{align}

From the supersymmetry transformations it is clear that the insertion of a single scalar field is BPS and one can check for consistency that
\begin{flalign}
 \bar Q Q C_a&=\{ Q, \bar Q\} C_a= -2\mathcal{D}_{\tau} C_a \qquad\qquad\, Q\bar Q C_{\dot a}=\{ Q, \bar Q\} C_{\dot a}=-2 \mathcal{D}_{\tau} C_{\dot a}\\
  \bar Q Q \bar C^{\dot a}&=\{ Q, \bar Q\} \bar C^{\dot a}= -2 \mathcal{D}_{\tau} \bar C^{\dot a} \qquad\qquad Q\bar Q \bar C^{ a}=\{ Q, \bar Q\} \bar C^{ a}=-2 \mathcal{D}_{\tau} \bar C^{ a}\,.
\end{flalign}

\subsection{Broken SUSY transformations}\label{broken}
In section \ref{dispmult} we need to compute the variation of the connection $\mathcal{L}_B$ under the broken generator $\mathfrak{Q}$. To do this we need to consider their action on the scalars and on the parallel component of the gauge field:
\begin{align}
 \mathfrak{Q} C_a&=0 & \mathfrak{Q} C_{\dot a}&= \e_{\dot a \dot b}\bar \psi^{\dot b}_+ & \mathfrak{Q} \bar C^a & =\e^{ab} \psi_b^- & \mathfrak{Q} \bar C^{\dot a} & =0
 \end{align}
 \begin{align}
 \mathfrak{Q} A_1&=\frac{2\pi i}{k}\left(\e^{ab} C_a \psi^-_b + \e_{\dot a \dot b} \bar \psi_+^{\dot a} \bar C^{\dot b} \right) \,.
\end{align}

Furthermore in section \ref{scalarmult} and \ref{Wardbrok} and we need the action of the supercharges $\mathfrak{Q}^{a\dot a}$ on the fundamental fields. For the scalars
\begin{align}
 \mathfrak{Q}^{a\dot a} C_b&=-  \d^a_b \bar \psi_+^{\dot a} & \mathfrak{Q}^{a\dot a} C_{\dot b}&= \d^{\dot a}_{\dot b} \bar \psi_+^{a} & \mathfrak{Q}^{a\dot a} \bar C^b&=-\e^{ab} \e^{\dot a \dot b} \psi^-_{\dot b} & \mathfrak{Q}^{a\dot a} \bar C^{\dot b}&= \e^{ab} \e^{\dot a \dot b} \psi_b^-
\end{align}
For the fermions
\begin{align}
 \mathfrak{Q}^{b\dot b} \psi^+_a&=-\d_a^b 2 \mathcal{D}_{\tau} \bar C^{\dot b}-\tfrac{8\pi }{k}[\d_a^b (\bar C^c C_c \bar C^{\dot b}-\bar C^{\dot b}C_c \bar C^c)+ \bar C^{\dot b} C_a \bar C^b -\bar C^b C_a \bar C^{\dot b}] \\
 \mathfrak{Q}^{b\dot b} \psi^-_a&=2 i\d_a^b D \bar C^{\dot b}\\
 \mathfrak{Q}^{b\dot b} \psi^+_{\dot a}&=\d_{\dot a}^{\dot b} 2  \mathcal{D}_{\tau} \bar C^{b}+\tfrac{8\pi }{k}[\d_{\dot a} ^{\dot b} (\bar C^c C_c \bar C^{ b}-\bar C^{ b}C_c \bar C^c) -\bar C^{\dot b} C_{\dot a} \bar C^{ b}+ \bar C^{ b} C_{\dot a} \bar C^{\dot b}]\\
 \mathfrak{Q}^{b\dot b} \psi^-_{\dot a}&=-2i \d_{\dot a}^{\dot b} D \bar C^{ b}\\
 \mathfrak{Q}^{b \dot b} \bar \psi_-^a&=\e^{ab} \e^{\dot a \dot b}2\mathcal{D}_{\tau} C_{\dot a}-\tfrac{8\pi}{k} [\e^{ab}\e^{\dot a \dot b} (C_{\dot c} \bar C^{\dot c} C_{\dot a}-C_{\dot a} \bar C^{\dot c} C_{\dot c})+\e^{cb} \e^{\dot c \dot b}(C_{\dot c} \bar C^a C_c- C_c \bar C^a C_{\dot c}) ] \\
 \mathfrak{Q}^{b \dot b} \bar \psi_+^a&=2i \e^{ab} \e^{\dot a \dot b} D C_{\dot a}\\
 \mathfrak{Q}^{b \dot b} \bar \psi_-^{\dot a}&=-\e^{ab} \e^{\dot a \dot b}2\mathcal{D}_{\tau} C_{ a}+\tfrac{8\pi}{k} [\e^{ab}\e^{\dot a \dot b} (C_{\dot c} \bar C^{\dot c} C_{ a}-C_{ a} \bar C^{\dot c} C_{\dot c})-\e^{cb} \e^{\dot c \dot b}(C_{\dot c} \bar C^{\dot a} C_c- C_c \bar C^{\dot a} C_{\dot c}) ] \\
 \mathfrak{Q}^{b \dot b} \bar \psi_+^{\dot a}&=-2 i\e^{ab} \e^{\dot a \dot b} D C_{a}
\end{align}
and for the parallel component of the gauge field
\begin{align}
 \mathfrak{Q}^{b\dot b} A_1 =-\frac{2\pi i}{k} (\e^{ab} \e^{\dot a \dot b} C_a \psi^-_{\dot a}-\e^{ab} \e^{\dot a \dot b} C_{\dot a} \psi^-_{ a}+\bar \psi^b_+ \bar C^{\dot b}-\bar \psi_+^{\dot b}\bar C^b)\,.
\end{align}
One particular supersymmetry transformation that we need in section \ref{Wardbrok} is
\begin{align}\label{QonF}
  \mathfrak{Q}^{a \dot a} \bar{\mathbb{F}}&=-2 \mathcal{D}_{\tau} \bar{\mathbb{O}}^{\dot a a}-\frac{4 \pi i}{k}\e^{\dot a \dot b} \bar \psi_{\a}^a \psi_{\dot b}^{\a}\\
  &-\tfrac{32\pi^2 i}{k^2} [\e^{\dot a \dot b}(C_{\dot c} \bar C^{\dot c} C_{\dot b}\bar{C}^a-C_{\dot b} \bar C^{\dot c} C_{\dot c}\bar C^a)+\e^{ac} \e^{\dot a \dot c}\e_{bd}(C_{\dot c} \bar C^b C_c\bar C^d- C_c \bar C^b C_{\dot c}\bar C^d) \nonumber \\
  &\phantom{-\tfrac{32\pi^2 i}{k^2}}-\e^{\dot a \dot c} (C_{\dot c}\bar C^c C_c \bar C^{ a}-C_{\dot c}\bar C^{ a}C_c \bar C^c) +\e^{\dot b \dot c} (C_{\dot c}\bar C^{\dot a} C_{\dot b} \bar C^{ a}-C_{\dot c} \bar C^{ a} C_{\dot b} \bar C^{\dot a})]\,.\nonumber
\end{align}

\bibliographystyle{nb}
\bibliography{biblio}

\end{document}